%% file: radiopurity_paper.tex
\RequirePackage{fix-cm}
\documentclass[twocolumn,epjc3]{svjour3}  
\smartqed  
\RequirePackage{graphicx,subfigure}
\RequirePackage{mathptmx}      
\RequirePackage{flushend}
\RequirePackage[numbers,sort&compress]{natbib}
\RequirePackage[colorlinks,citecolor=blue,urlcolor=blue,linkcolor=blue]{hyperref}
\RequirePackage{amsmath,amssymb}
\RequirePackage{units}
\usepackage{pdflscape}
\usepackage{afterpage}
\usepackage{capt-of}
\usepackage{microtype}
\usepackage[mathlines]{lineno}
\usepackage{color}
\usepackage{isotope}   
\usepackage[separate-uncertainty=true]{siunitx} 
\usepackage{scrextend}
\usepackage{lipsum}
\usepackage{rotating}
\usepackage{textcomp}
\usepackage[dvipsnames]{xcolor}
\usepackage[british]{babel}

\journalname{Eur. Phys. J. C}

\renewcommand{\figurename}{Figure}

\usepackage{titlesec}
\titleformat{\paragraph}
{\normalfont\normalsize\bfseries}{\theparagraph}{1em}{}
\titlespacing*{\paragraph}
{0pt}{3.25ex plus 1ex minus .2ex}{1.5ex plus .2ex}

\begin{document}
\def\rnAccuglassa{\hyperref[tab:rn_items]{Rn1}}
\def\rnAccuglassb{\hyperref[tab:rn_items]{Rn2}}
\def\rnAccuglassc{\hyperref[tab:rn_items]{Rn3}}
\def\rnHabia{\hyperref[tab:rn_items]{Rn4}}
\def\rnHubersuhnera{\hyperref[tab:rn_items]{Rn5}}
\def\rnHubersuhnerb{\hyperref[tab:rn_items]{Rn6}}
\def\rnHubersuhnerc{\hyperref[tab:rn_items]{Rn7}}
\def\rnPasternack{\hyperref[tab:rn_items]{Rn8}}
\def\rnUhmwpe{\hyperref[tab:rn_items]{Rn9}}
\def\rnIglidurwhite{\hyperref[tab:rn_items]{Rn10}}
\def\rnIgliduryellow{\hyperref[tab:rn_items]{Rn11}}
\def\rnIglidurblue{\hyperref[tab:rn_items]{Rn12}}
\def\rnViton{\hyperref[tab:rn_items]{Rn13}}
\def\rnSSvacuumline{\hyperref[tab:rn_items]{Rn14}}
\def\rnSScryovalve{\hyperref[tab:rn_items]{Rn15}}
\def\rnAlrotor{\hyperref[tab:rn_items]{Rn16}}
\def\rnXenTCryostat{\hyperref[tab:rn_XENONnT_samples]{Rn19}}
\def\rnXenTCryogenics{\hyperref[tab:rn_XENONnT_samples]{Rn20}}
\def\rnXenTNewCablePipe{\hyperref[tab:rn_XENONnT_samples]{Rn22}}
\def\rnXenTNewCableFeedthrough{\hyperref[tab:rn_XENONnT_samples]{Rn21}}
\def\rnXenTGXeMaga{\#23}
\def\rnXenTGXeGettera{\#16}
\def\rnXenTGXeGetterb{\#17}
\def\rnXenTLXePumpb{\hyperref[tab:rn_XENONnT_samples]{Rn27}}
\def\rnXenTLXePumpabefore{A28$_{b}$}
\def\rnXenTLXeVacumPipe{\hyperref[tab:rn_XENONnT_samples]{Rn23}}
\def\rnXenTLXePurityMonitor{\hyperref[tab:rn_XENONnT_samples]{Rn24}}
\def\rnXenTLXeHeatExchanger{\hyperref[tab:rn_XENONnT_samples]{Rn26}}
\def\rnXenTLXeFilterQa{\hyperref[tab:rn_items]{Rn17}}
\def\rnXenTLXeFilterQb{\hyperref[tab:rn_items]{Rn18}}
\def\rnXenTLXeFilterST{\hyperref[tab:rn_XENONnT_samples]{Rn25}}
\def\rnXenTtpc{\hyperref[tab:rn_XENONnT_samples]{Rn35}}
\def\rnXenTRADSecespol{\hyperref[tab:rn_XENONnT_samples]{Rn33}}
\def\rnXenTRADMagcdef{\hyperref[tab:rn_XENONnT_samples]{Rn30}}
\def\rnXenTRADreboilerdown{\hyperref[tab:rn_XENONnT_samples]{Rn34}}
\def\rnXenTRADbufferin{\hyperref[tab:rn_XENONnT_samples]{Rn31}}
\def\rnXenTRADbufferout{\hyperref[tab:rn_XENONnT_samples]{Rn32}}
\def\rnXeoneTRADgetter{\hyperref[tab:rn_XENONnT_samples]{Rn29}}
\def\rnXenTRADColumn{\hyperref[tab:rn_XENONnT_samples]{Rn28}}
\def\rnXenTCryoIntegral{21\,(3)}
\def\rnXenTGXeIntegral{1.7\,(2)}
\def\rnXenTLXeIntegral{3.6\,(2)}
\def\rnXenTRnDSTIntegral{1.6\,(2)}
\def\rnXenTtpcIntegralStat{9.3\,(3.8)}
\def\rnXenTtpcIntegral{9.3\,(3.8){\footnotesize \emph{stat}}\,($^{+1.2}_{-4.6}$){\footnotesize \emph{sys}}}
\def\rnXenTTotal{35.7\,($^{+4.5}_{-5.9}$)}
\def\rnXenTInventory{8.4}
\def\rnXenTConcentration{4.2\,($^{+0.5}_{-0.7}$)}

\title{Material radiopurity control in the XENONnT experiment
}

\input{authorlist_20210921.tex}





\date{\phantom{Received: date / Accepted: date}}

\maketitle

\begin{abstract}
The selection of low-radioactive construction materials is of the utmost importance for rare-event searches and thus critical to the XENONnT experiment.
Results of an extensive radioassay program are reported, in which material samples have been screened with gamma-ray spectroscopy, mass spectrometry, and $^{222}$Rn emanation measurements.
Furthermore, the cleanliness procedures applied to remove or mitigate surface contamination of detector materials are described.
Screening results, used as inputs for a XENONnT Monte Carlo simulation, predict a reduction of materials background ($\sim$17\%) with respect to its predecessor \mbox{XENON1T}.
Through radon emanation measurements, the expected $^{222}$Rn activity concentration in XENONnT is determined to be \rnXenTConcentration\,{\textmu}Bq/kg, a factor three lower with respect to \mbox{XENON1T}.
This radon concentration will be further suppressed by means of the novel radon distillation system.

\keywords{Gamma spectroscopy \and Mass spectrometry \and Radon emanation \and Surface treatment\and Detector cleanliness}
\end{abstract}

\section{Introduction}
The XENONnT detector was constructed for the direct detection of weakly interacting massive particles (WIMPs)~\cite{ref:wimp}, a widely discussed dark matter candidate. Additionally, due to the low background levels achieved, it will contribute to a wide array of other rare event searches, such as the two-neutrino double electron capture in $^{124}$Xe \cite{ref:xenon1t_dec}, neutrino-less double-beta decay of $^{136}$Xe \cite{ref:dbb_xe136}, solar-axions \cite{ref:er_excess}, and coherent elastic scattering of solar neutrinos \cite{ref:b8_neutrinos}.
The detector, located in the underground Laboratori Nazionali del Gran Sasso (LNGS), operates as a dual-phase time projection chamber (TPC) with a 5.9\,tonnes liquid xenon (LXe) target. 
Incident particles are observed either through scattering off a xenon nucleus or its electron cloud, which are known as nuclear recoils (NR) and electronic recoils (ER), respectively.
\mbox{XENONnT} aims to probe spin-independent WIMP-nucleon cross sections down to $1.4\cdot 10^{-48}\,$cm$^2$
for a 50\,GeV/c$^{2}$ WIMP at 90\,\% confidence level (C.L.)~\cite{ref:xenonnt_mc}.

To reach the low background requirements for \mbox{XENONnT}, special focus is set on the background mitigation. In this context, a careful selection of the detector materials plays a key role in the suppression of trace radioactive contaminants that contribute to the overall background \cite{ref:screening_lz, ref:screening_panda, ref:xenon1t_gamma, ref:xenon1t_radon}. 
Detrimental sources of background are the primordial isotopes $^{232}$Th, $^{238}$U, $^{235}$U, $^{40}$K and their progeny, as well as $^{60}$Co and $^{137}$Cs.
These contaminants may be inherent to the raw material or get introduced during the production of detector components. Gamma-ray spectroscopy and mass spectrometry are employed to determine the intrinsic radioactivity of construction materials and can reach sensitivities down to \mbox{10--100}\,{\textmu}Bq/kg and \mbox{1--10}\,{\textmu}Bq/kg, respectively.

Another important selection criterion for materials is the emanation rate of $^{222}$Rn. Produced in the decays of residual $^{226}$Ra, which is present in nearly all materials, the radioactive noble gas $^{222}$Rn might be released into the LXe target. There, its progeny can induce low-energy ER background events throughout the sensitive volume. The material selection of \mbox{XENONnT} was aimed to achieve a $^{222}$Rn activity concentration of 1\,{\textmu}Bq/kg in the LXe target during standard operation~\cite{ref:xenonnt_mc}.

The long-lived radon daughter $^{210}$Pb and its progeny can also plate-out on material surfaces, mostly before the detector assembly, and thus contribute to the overall background. While the beta decays of $^{210}$Pb and $^{210}$Bi contribute to the so-called 
`surface background', as observed in \mbox{XENON1T} \cite{ref:xenon1t_analysis2}, the alpha-decay of $^{210}$Po can enhance the neutron-induced background through ($\alpha$,n) reactions \cite{ref:xenonnt_mc,ref:lz_mcsim}. A dedicated material surface treatment procedure in combination with cleanroom facilities optimized for storage and detector assembly were developed in order to mitigate radioactive surface contamination.

This paper describes the radiopurity measures applied during the construction of the \mbox{XENONnT} experiment. 
In Section~\ref{sec:2}, a brief detector introduction is given. A more detailed description can be found in~\cite{ref:xenonnt_mc}.
Section~\ref{sec:gamma} summarizes the results obtained in the gamma screening campaign including complementary mass spectrometry measurements. The radon emanation measurements campaign is described in Section~\ref{sec:radon}, while Section~\ref{sec:clean} summarizes the cleanliness measures applied during the detector assembly.
The results are summarized in Section~\ref{sec:results}, where the expected background rates, estimated based on the output of the described screening efforts, are discussed.

\section{The XENONnT Experiment}
\label{sec:2}
\begin{figure*}[t!]
\centering 
\includegraphics[width=0.95\textwidth]{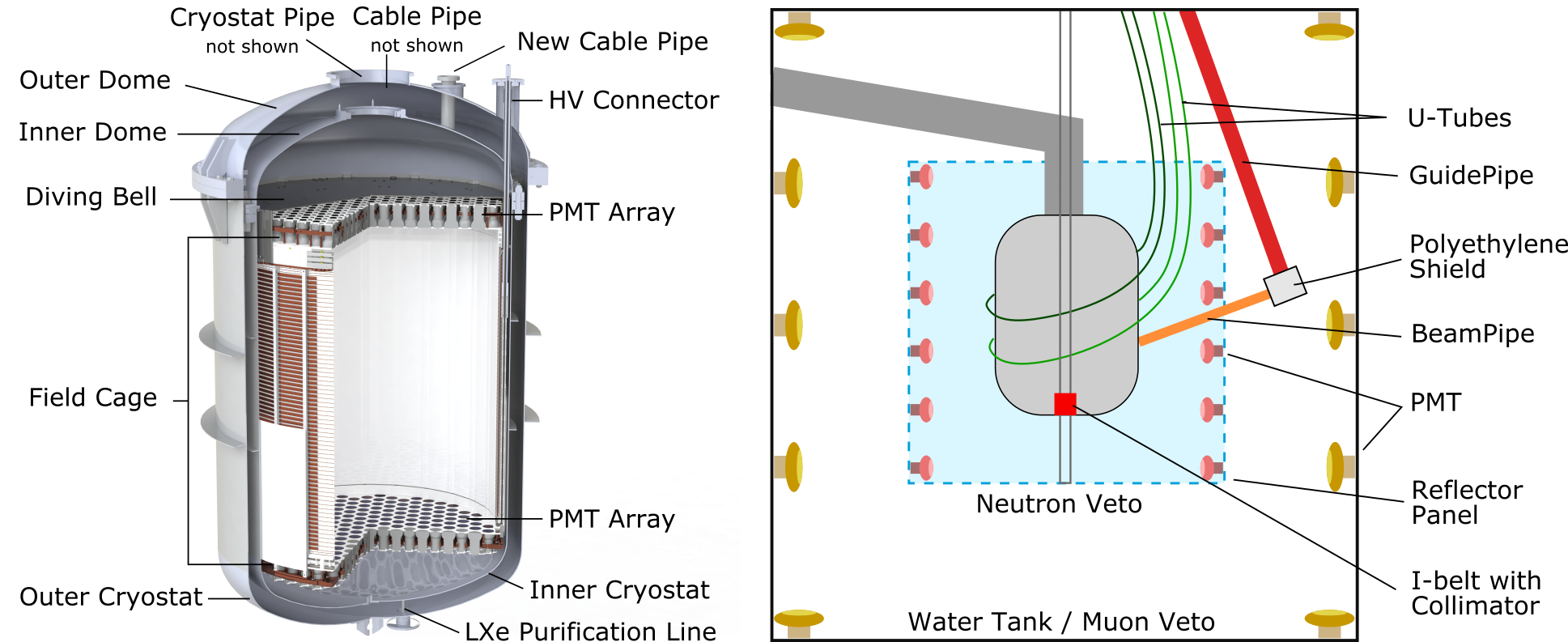} 

\caption{(left) Render of the XENONnT cryostat and TPC, including the most relevant components investigated during the screening campaign. 
%
(right) The cryostat is surrounded by the neutron veto and muon veto which serve as shielding for background reduction. The U-Tubes, Guide Pipe, Beam Pipe, and I-belt are part of the calibration subsystem.}
\label{fig:sketch_cryostat}
\end{figure*}
The \mbox{XENONnT} TPC has a diameter of 1.3\,m and a height of 1.5\,m. It is housed inside a double-walled vacuum-insulated cryostat vessel, as illustrated in Figure~\ref{fig:sketch_cryostat} (left). 
The TPC is composed of a field cage, electrodes, reflector panels, support pillars,
and 494 photomultiplier tubes (PMTs) arranged in two arrays, one each at the top and bottom of the TPC region.
Polytetrafluoroethylene (PTFE) reflector panels at the wall 
provide high reflectivity minimizing the loss of primary scintillation light created in particle interactions in the LXe target.
Signal and high voltage (HV) cables of the PMTs are guided to the outside through two cable pipes that are connected to the cable feedthroughs.
Vertical PTFE pillars serve as the frame for installing the reflector panels, guard rings, and field-shaping rings. 
The latter two are made from oxygen-free high conductivity (OFHC) copper and ensure the homogeneity of the electric field needed to drift free electrons toward the liquid-gas interface at the top of the TPC. Once at the interface, a stronger electric field extracts them into the gaseous xenon (GXe) region to create a secondary scintillation signal.
The stainless steel (SS) frames of the electrodes (e.g., cathode and anode) hold the parallel electrode wires, with each electrode kept at different potential. 
The entire TPC hangs from the diving bell, which is immersed in LXe. 
The LXe level inside the TPC is controlled by pressurizing the diving bell with GXe. 
The outer insulation cryostat vessel and its flange (outer dome), the functional PMTs, and cables were reused from \mbox{XENON1T}~\cite{ref:xenon1t_instruments}.

\begin{figure*}[h]
\centering 
\includegraphics[width=0.9\textwidth]{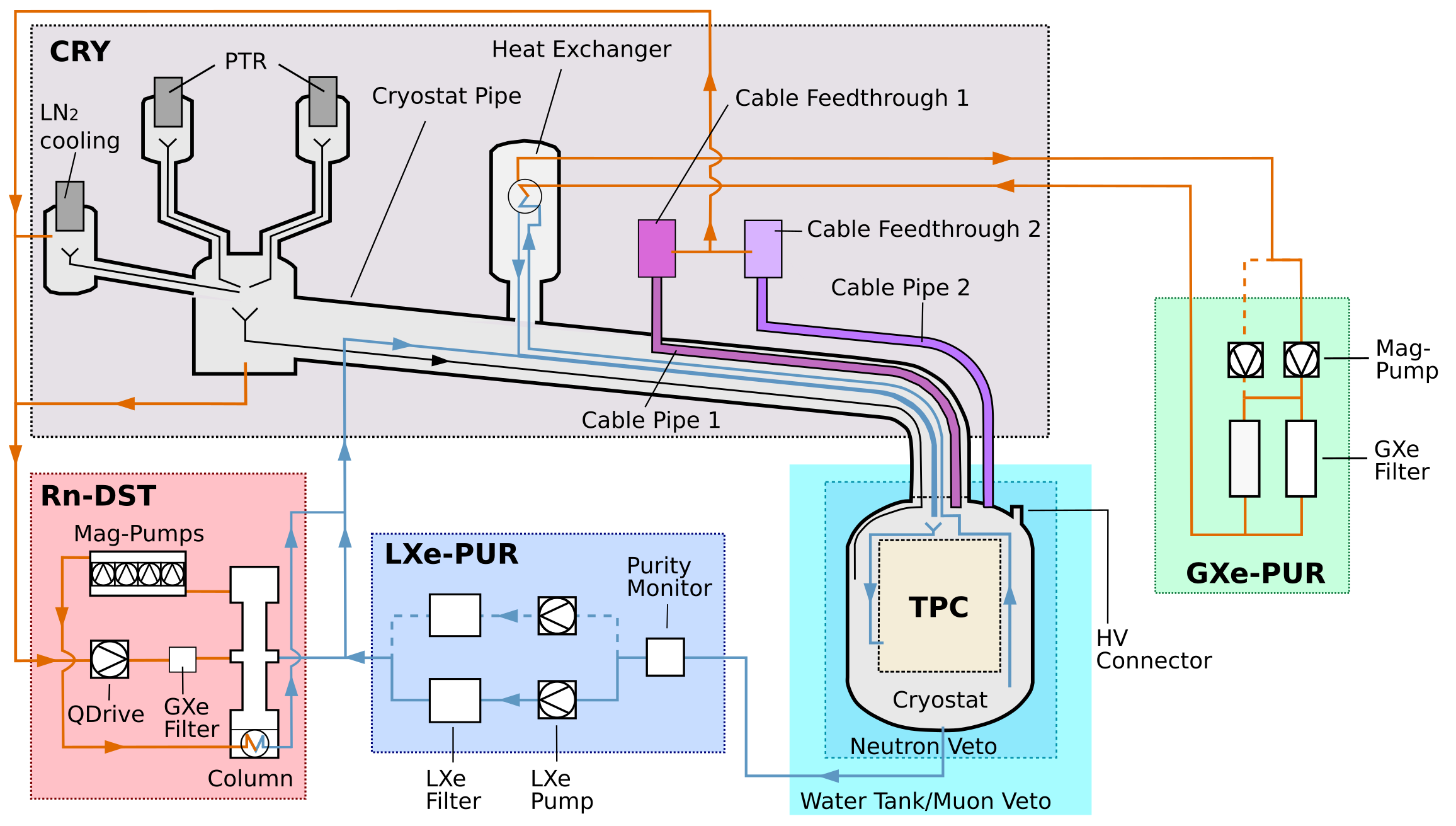}
\caption{Schematic drawing of the xenon handling system and the TPC of XENONnT. 
The circulation of xenon through the purification systems is indicated with orange (GXe) and blue (LXe) flow lines.\vspace{1.0cm}
}
\label{fig:sketch_xenonnt}
\end{figure*}

For operating \mbox{XENONnT}, a total xenon inventory of \rnXenTInventory\,tonnes is needed, including 5.9\,tonnes of active target mass inside the TPC.
The schematics of the 
xenon-handling system
are shown in Figure~\ref{fig:sketch_xenonnt} (the diving bell has been omitted for better visualization).
One major component is the cryogenic system (CRY).
It includes two pulse tube refrigerators (PTR)
and a liquid nitrogen (LN$_2$) emergency cooling system, which have been retained from \mbox{XENON1T}~\cite{ref:xenon1t_instruments}. 
The Cryostat Pipe, also entirely reused from \mbox{XENON1T}, connects the cryostat to the rest of the CRY system and houses Cable Pipe~1 and other piping used for xenon purification and liquefaction. 
The Cable Pipe~2 is placed outside of the Cryostat Pipe as it was installed newly during the upgrade to
\mbox{XENONnT} to accommodate additional cabling due to the increased number of PMTs.

Two purification systems constantly clean the LXe target of electronegative impurities. In the GXe purification (\mbox{GXe-PUR}) system, LXe from the cryostat is evaporated and circulated through two high-temperature rare-gas purifiers operated in parallel (GXe Filter in Figure~\ref{fig:sketch_xenonnt}).
A magnetically coupled piston pump (Mag-Pump)~\cite{ref:muenster_magpump} achieves purification flows of $\sim$50\, standard liters per minute (slpm).
The same pump is also procuring the GXe to pressurize the diving bell. 
A second spare Mag-Pump is also available for redundancy.

The new LXe purification (\mbox{LXe-PUR}) system is operated parallel to the \mbox{GXe-PUR} and enables much higher purification flows up to 3 LXe liters per minute (equivalent to 1500 slpm).
LXe is extracted from the bottom of the cryostat and flows through vacuum-insulated pipes into the purification unit. 
A LXe pump (model BNCP-32-000\footnote{www.barber-nichols.com}) circulates LXe through a rare-gas purifier material that binds trace amounts of electronegative impurities in the LXe (LXe filter in Figure~\ref{fig:sketch_xenonnt}, using the same reactive material as the GXe filter). 
Since the LXe filter needs to be regenerated after saturation, 
the \mbox{LXe-PUR} possesses a redundant LXe pump and filter unit to guarantee continuous operation during the regeneration process.
The \mbox{LXe-PUR} is also instrumented with a purity monitor module, which is able to measure the level of electronegative impurities in the LXe coming from the cryostat. 

\mbox{XENONnT} is the first experiment to have a dedicated online radon removal system based on cryogenic distillation \cite{ref:mpik_radon_distillation,ref:xenon100_radon_distillation}. 
While the majority of the purified xenon from the \mbox{LXe-PUR} is returned directly to the cryostat, up to 200\,slpm of LXe are directed to the radon distillation column (\mbox{Rn-DST}) to remove trace amounts of radon. Due to the lower vapor pressure of radon compared to xenon, radon accumulates in the LXe at the bottom of the column while the GXe extracted at the top is radon depleted. 
Four Mag-Pumps~\cite{ref:4magpumps}, similar to the ones used in the \mbox{GXe-PUR}, compress the radon-depleted GXe inside a heat-exchanger which is in thermal contact with the column's LXe reservoir.
There, the radon-depleted xenon is liquefied and returned to the cryostat. 
In addition to the radon-distillation of LXe coming from the \mbox{LXe-PUR}, there is the possibility to 
distill GXe extracted from different detector locations. 
By doing so, radon that is emanated in detector subsystems, such as the cable pipes, gets removed before entering the TPC's active region. 
The necessary GXe recirculation flow is achieved with
a customized QDrive piston pump from Chart Industries\footnote{https://www.chartindustries.com/} retained from \mbox{XENON1T}.
Electronegative impurities from the GXe are also removed by circulating through a GXe filter unit similar to the ones in \mbox{GXe-PUR}.

For shielding and background mitigation, the cryostat is located inside a water tank, 
which also functions as an active muon veto and neutron veto (as seen in Figure~\ref{fig:sketch_cryostat}, right).
The water tank and the Cherenkov muon veto system are retained from the \mbox{XENON1T} experiment~\cite{ref:xenon1t_instruments,ref:muon_veto}. The newly-built neutron veto system encloses the region around the cryostat with light reflective panels and is operated with 120 PMTs. 
Gadolinium-sulfate, which will be dissolved in the entire water volume, efficiently captures neutrons leaving the cryostat~\cite{ref:gd_kamiokande}. As a consequence of the capturing process, Cherenkov light is produced which is detected.

In addition to the internal calibration
sources (e.g., $^{220}$Rn and $^{83\text{m}}$Kr) which are periodically flushed directly into the LXe target, external sources are also used. 
Two U-tubes extend from the top of the water tank and encircle the cryostat which allow the deployment of external sources close to the detector.
Moreover, an I-belt system can be used to move a payload vertically along the cryostat, similar to \mbox{XENON1T}~\cite{ref:xenon1t_instruments}. The I-belt can carry a tungsten collimator containing radioactive sources, e.g. YBe for low-energy neutrons.

Other neutron sources may also be lowered from the top of the water tank inside the Guide Pipe from which collimated neutrons reach the cryostat and TPC through the Beam Pipe. A boronated polyethylene shield surrounds the bottom of the Guide Pipe to reduce neutron captures in the water tank while a neutron source is deployed.

The muon veto, neutron veto, and the calibration systems are not in direct contact with the xenon (see Figure~\ref{fig:sketch_cryostat} right) and contribute subdominantly to the overall level of background for dark matter searches.
However, since they do contribute to the n-veto background and consequently impact its efficiency, also the materials used for the n-veto and the new calibration system have been screened and carefully selected.

\section{Radioassay Program}
\label{sec:gamma}
The techniques of gamma-ray spectroscopy and mass spectrometry were employed to provide information about the specific activities of radionuclides in detector materials. 
Acceptable radio-isotopic concentrations for any given batch of raw material vary according to the component's mass and its proximity to the active region of the TPC.
Gamma-ray spectroscopy is a non-invasive method sensitive to a wide range of gamma emitters. Samples with masses ranging from a few grams up to 100\,kg were measured with a high-purity germanium crystal for \mbox{10--40}\,days to reach the sensitivity that current subterranean experiments require. 
Inductively Coupled Plasma Mass Spectrometry (\mbox{ICP-MS})~\cite{ref:ICPMSnisi,ref:ICPMSvacri} can determine the composition of a sample by separating and measuring individual isotopes, such as $^{238}$U and $^{232}$Th. The higher sensitivity, as well as the smaller sample sizes and measuring times needed for \mbox{ICP-MS}, make it a complementary method to gamma-ray spectroscopy.

The XENON collaboration employs several highly sensitive germanium spectrometers: Gator~\cite{ref:Gator}, GeMSE~\cite{ref:GeMSE}, and four GeMPI spectrometers~\cite{ref:GeMPI}.
The Gator facility and GeMPI detectors sit near \mbox{XENONnT} at LNGS, while GeMSE is located in the Vue-des-Alpes underground laboratory in Switzerland. Each spectrometer is an intrinsically pure p-type germanium crystal (HPGe) set in a coaxial configuration and housed in a low-radioactivity cryostat. The crystal, with a mass of $\sim$\mbox{2--3}\,kg, extends into an inner active region enclosed by OFHC copper. The inner chambers are continuously purged with gaseous nitrogen to counteract the influx of ambient radon. The copper is surrounded by \mbox{20--25}\,cm of lead, where the innermost layer has the lowest levels of $^{210}$Pb contamination. These detectors can reach sensitivities of $\sim$10\,{\textmu}Bq/kg. GSOr, GeCris and GeDSG, which are part of the SubTErranean Low Level Assay (STELLA) laboratory~\cite{ref:stella} at LNGS, were also used when needed. They have sensitivities of \mbox{1--10}\,mBq/kg.

Three additional p-type HPGe facilities, Bruno, Corrado, and GIOVE~\cite{ref:Giove} \mbox{0.9--1.8}\,kg, were utilized for smaller components and cleaning agents. These detectors were operated in the underground Low-Level Laboratory at Max Planck Institut f\"{u}r Kernphysik in Heidelberg. These spectrometers are shielded by copper and lead,
and are equipped with an active muon veto. 
In the case of GIOVE, neutron background is further reduced with an additional borated Polyethylene based shielding. 
These facilities can reach sensitivities of \mbox{0.1--1}\,mBq/kg.

Samples screened in the HPGe facilities were cleaned with mildly acidic soap (e.g. Elma clean 65), rinsed with deionized water (DI water), and immersed in ethanol ($>$95\%). Both steps were completed with a 20-minute ultrasonic bath (US-bath). If a sample could not be cleaned with acidic soap or immersed in liquid (e.g. photomultipliers and cables), its surface was wiped thoroughly with ethanol. All samples were stored within clean plastic bags to mitigate plate-out of radon daughters during transport. Prior to the measurement it was verified that all traces of ambient radon and radon daughters have been removed or decayed by monitoring the count rates of the associated gamma lines.

The Geant4 toolkit~\cite{ref:geant4} was used to simulate each individual sample inside the respective HPGe spectrometer in order to ascertain the detection efficiency for each gamma line. 
The specific activities (or upper limits) were then calculated from data based on the sample's mass and measuring time, as well as the characteristic branching ratios of the gamma lines, as detailed in~\cite{ref:Gator}.

The complementary analytic technique \mbox{ICP-MS} is among the most sensitive for the detection of trace elements. The intrinsic radioactivity of a batch of material can be found with a measurement of long-lived radionuclides. The sample is turned into an aqueous solution, introduced through a peristaltic pump, nebulized in a spray chamber, and then atomized and ionized in plasma. These ions are extracted into a system placed under high vacuum and separated in accordance with the charge-to-mass ratio. 
Sensitivities \mbox{1--10}\,{\textmu}Bq/kg 
are attainable for $^{238}$U and $^{232}$Th. 

Several live years of data, aggregated across all instruments, were acquired throughout the radioassay program of \mbox{XENONnT}. Relevant measurements for detector construction are covered here, where more supplemental data is available in~\cite{ref:gamma_zenodo}. 
For detected activity, the 1$\sigma$ uncertainties are given, including both statistical and systematic contributions. Systematic uncertainties result primarily from efficiency simulations. Otherwise, upper limits are provided at 95\,\%~C.L. In the case of \mbox{ICP-MS}, uncertainties are given to account for instrumental precision, calibration, and the recovery efficiency. 
A break in secular equilibrium in the $^{238}$U decay chain is identified by comparing the results obtained for $^{238}$U and $^{226}$Ra from gamma spectroscopy. A deviation from the $^{228}$Ra result from the direct measurement of $^{232}$Th with \mbox{ICP-MS} indicates a break in secular equilibrium in the $^{232}$Th chain.
The results of materials and cleaning agents (discussed in Section~\ref{sec:cr_infrastructure}) selected for use in the \mbox{XENONnT} TPC and cryostat are shown in Table~\ref{tab:screening}. 
Components fabricated from these materials are displayed in Figure~\ref{fig:sketch_cryostat} (left). 
These components contribute substantially to the overall background rate of \mbox{XENONnT}. These results are incorporated into the sensitivity study of \mbox{XENONnT}~\cite{ref:xenonnt_mc} through a Monte Carlo simulation of the material-induced background. Whenever only an upper limit is available, the upper limit is assumed as the activity to acquire the most conservative sensitivity.

Similarly, Table~\ref{tab:screening_other} lists the specific activities of components selected for usage in the neutron veto, calibration and purification subsystems. These components do not contribute significantly to the materials-induced background for the dark matter search, but they determine the acceptable tagging window and coincidence threshold for the PMTs of the neutron veto system.

Overall, the inherent concentrations of isotopic impurities in the bulk materials used in \mbox{XENONnT} are comparable to the materials used in \mbox{XENON1T}~\cite{ref:xenon1t_gamma}. Some of the PTFE (Sample 9) from which TPC wall reflectors were fabricated proved to be higher than expected in $^{40}$K but lower in $^{238}$U, $^{226}$Ra and $^{137}$Cs. The remaining wall reflectors (Sample 10), on the other hand, were made of material low in $^{40}$K and $^{137}$Cs, but higher in $^{226}$Ra. The SS material selected in \mbox{XENONnT} for the electrode frames, bell, and inner cryostat vessel, came after several batches were rejected because they would have contributed substantially to the NR background through spontaneous fission and ($\alpha$,n) reactions. The selected SS material also exhibited low specific activities of $^{60}$Co ($\sim$1~mBq/kg), which is the most substantial contributor to the ER background in that material.
Much of the dedicated efforts were additionally focused on the individual components of the PMTs. Most of the components showed similar levels of impurities to those in \mbox{XENON1T}~\cite{ref:xenon1t_pmt}. Results for the ceramic stems are given in Table~\ref{tab:screening} as they are the largest contributor to the total radioactivity of the PMTs.
OFHC copper did not exhibit any deviation from the expected purity. Multiple earlier samples of Gadolinium Sulfate proved to have higher concentrations of all isotopes, except $^{137}$Cs, by \mbox{1--2} orders of magnitude. Table~\ref{tab:screening_rejected} lists the results for example materials that were rejected in the course of the radioassay program. A comparison of spectroscopic and spectrometric results shows no significant break in secular equilibrium in the $^{232}$Th chain for any screened material.

\section{$^{222}$Rn emanation measurements}
\label{sec:radon}
The emanation rate of $^{222}$Rn from detector components cannot be inferred from gamma-screening due to the often unknown radon diffusion in materials and the potential inhomogeneous distribution of the mother isotope $^{226}$Ra. 
Thus, prior to the construction of XENONnT, all detector materials were investigated for their radon emanation. 
Furthermore, fully assembled
detector subsystems were measured for radon emanation 
in order to get a complete understanding of the locations of radon sources in the system. This information is needed to optimize the performance of the radon distillation system.

The results in this section refer to the $^{222}$Rn activity at its emanation equilibrium
and are given with a combined uncertainty including statistical and systematic errors, unless specified otherwise. If a result is compatible with zero within 1.645\,$\sigma$, a 90\,\% C.L. upper limit is given instead\footnote{Note that upper limits in Section~\ref{sec:gamma}, were given at 95\% C.L., in order to be consistent with previous \mbox{XENON1T} gamma-screening~\cite{ref:xenon1t_gamma} and radon-emanation~\cite{ref:xenon1t_radon} references.}.

\subsection{$^{222}$Rn assay technique}
\label{sec:222Rn_procedure}
The applied $^{222}$Rn assay techniques are described in detail in~\cite{ref:xenon1t_radon}. 
The investigated sample was left for several days inside a \mbox{gas-tight} vessel filled with a \mbox{radon-free} carrier gas at ambient temperature. Emanated $^{222}$Rn atoms from the sample accumulated in the carrier gas during this emanation time. Then, the carrier gas was pumped through a LN$_{2}$-cooled adsorbent trap, the so-called radon trap, where the radon was collected and separated from the carrier gas.
In case of an equilibrium between radon emanation and its decay, the activity of the trapped radon corresponds to the sample's emanation rate. It was measured using miniaturized proportional counters which reach sensitivities down to $\sim\,$20\,{\textmu}Bq~\cite{ref:galex_prop_counters,ref:xenon1t_radon}.
If the sample had been exposed to xenon prior to the measurement (e.g., all reused \mbox{XENON1T} systems), large xenon outgassing rates prevented the usage of the proportional counters, as the xenon gets collected in the radon trap as well. As a consequence, the proportional counter's active volume of $\sim\,$1\,cm$^{3}$ was too small to house the entire sample.

In that case, electrostatic radon monitors were used which have significantly larger volumes ($\sim\,$10$^{3}$~cm$^{3}$) and sensitivities of $\sim\,$0.1\,mBq~\cite{ref:radon_monitor,ref:phd_bruenner}.
The radon monitor does not detect the $^{222}$Rn decay directly, but its \mbox{alpha-decaying} daughters $^{218}$Po and $^{214}$Po which, due to an electric drift field, are collected on a $\alpha$-sensitive photodiode.
The detection efficiency for each daughter isotope can be different, but they both strongly depend on the gas composition. Outgassing impurities released from samples have been shown to impact the detection efficiency of radon monitors~\cite{ref:rd_monitor_impurities}. For this reason, after each measurement a calibrated amount of radon was added to the sample for a calibration in the present gas composition.

Xenon outgassing may also hinder the extraction of the carrier gas as it freezes inside the radon trap and blocks the gas flow.
Such an effect was already observed during the \mbox{XENON1T} radon screening campaign~\cite{ref:xenon1t_radon}. 
Therefore, the $^{222}$Rn collection was done in a 2-stage approach: Before the radon trap another trap was included, the so-called xenon trap. It is a SS vessel filled with copper wool and held at LN2 temperature during the extraction. Xenon and the majority of radon freezes out in this trap however, due to the loose packing of the copper wool and the relatively large cross section of the trap, the carrier gas flow hardly degrades.
All radon (and also xenon), which cannot be stopped in the xenon trap is collected in the subsequent radon trap which is also held at LN2 temperature. At the end of the extraction, the xenon trap is warmed and its entire content is transferred into the radon trap. 
With this extra step, the entire procedure doesn't suffer from any flow degradation due to a blocked trap and all extracted radon from the sample can be stored in the radon trap (as verified using a calibrated radon source). 
In case of large gas samples to be extracted, the pumping power through the radon trap was too weak to extract the entire sample. Then, the quoted activities are corrected for this reduced extraction efficiency assuming that the emanated radon was homogeneously distributed within the carrier gas prior to extraction. This correction is referred to as scaling. In order to generate a homogeneous radon distribution in large volume samples, the carrier gas was mixed prior to the sample extraction by adding additional clean carrier gas via multiple filling ports.

\subsection{Radon emanation of construction materials}
\label{sec:222Rn_items}
All the results of this section can be found in Table~\ref{tab:rn_items}.
Almost 14\,km of PMT signal \mbox{read-out} and HV cables run from the TPC to the cable feedthrough, through the CRY system. 
Several samples of HV cable from the company \mbox{Accu-Glass} were measured, all of them with emanation rates well-within the requirements (\rnAccuglassa, \rnAccuglassb\, and \rnAccuglassc). 
For the signal cables, initially a PTFE-insulated coaxial cable from HABIA was considered (item \rnHabia, same company and type as item \#42 in~\cite{ref:xenon1t_radon}), but
it was found to emanate a factor $\sim\,$50 more than the sample reported in~\cite{ref:xenon1t_radon}.
Alternatives from two different companies were explored instead:
three signal cable samples from the company Huber+Suhner (\rnHubersuhnera, \rnHubersuhnerb\, and \rnHubersuhnerc) and one from the company Pasternack (\rnPasternack). 
All of them gave similar results, well-within the requirements.
The cable's emanation results given above serve as an upper limit for the expected emanation of all
cables (HV and signal cables) enclosed inside Cable Pipe~2 and the cryostat (the emanation of cables inside Cable Pipe~1 was already studied in~\cite{ref:xenon1t_radon}).
A measurement of their final activities will be presented later.

The inner pistons of all \mbox{Mag-Pumps} are sealed from their outer cylinders by a plastic-type gasket.
An ultra high molecular weight polyethylene (\mbox{ULHWP}, used for the \mbox{Mag-Pump} in \mbox{XENON1T}~\cite{ref:muenster_magpump}) was measured (\rnUhmwpe).
Several alternative gasket materials were explored (\rnIglidurwhite, \rnIgliduryellow\,and \rnIglidurblue).
All considered options showed negligible contributions with respect to the expected overall radon emanation of the pump (based on sample \#23 in~\cite{ref:xenon1t_radon}).

The radon emanation from individual components from the two LXe pumps was measured separately.
Results are shown as \rnViton-\rnAlrotor\,. 
The SS cryogenic valve and aluminum rotor were found to contribute negligibly while Viton O-rings might contribute notably to the total emanation of a LXe pump. 
Here it should be noticed that only a fraction of the \mbox{O-ring's} surface is expected to emanate radon into the LXe volume.

For the oxygen removal in the \mbox{LXe-PUR} system, 
a filter material made from copper electrolytically deposited onto alumina balls was considered.
Different batch samples of the same filter material were investigated for their radon emanation (samples \rnXenTLXeFilterQa\,and \rnXenTLXeFilterQb).
They also appear in Table~\ref{tab:screening}, 
samples 48-50, where the differences observed in $^{222}$Rn emanation rates can be related to the $^{226}$Ra activities.
While this filter has a very high oxygen removal rate, its radon emanation is at least a factor of 100 larger than the eventually used LXe filter discussed below.


\subsection{XENONnT radon emanation activity}
\label{sec:222Rn_samples}
The radon emanation rate of all detector subsystems introduced in Figure~\ref{fig:sketch_xenonnt} was measured separately. The results are listed in Table~\ref{tab:rn_XENONnT_samples}. Based on these measurements, the locations of all relevant radon sources were identified.
Some of the samples, or even entire subsystems of \mbox{XENONnT}, were already measured in preparation of the \mbox{XENON1T} experiment. 
Their emanation results reported in~\cite{ref:xenon1t_radon} are also included here.

An important result is the emanation from the SS inner cryostat (\rnXenTCryostat), where only surfaces facing inwards contributed to the measurement. 
Despite having a five times larger surface area, the 
emanation rate is comparable with the one obtained for the \mbox{XENON1T} inner cryostat (sample \#48 in~\cite{ref:xenon1t_radon}). 
The \mbox{XENON1T} cryogenic system was reused for \mbox{XENONnT}, and was kept under a nitrogen atmosphere during the detector upgrade. 
The result of an integral measurement (\rnXenTCryogenics) of the Cable Feedthrough~1, Cable Pipe~1 and the reused cooling towers of \mbox{XENON1T} was consistent with the emanation rate measured in~\cite{ref:xenon1t_radon} (obtained by adding together samples \#13, \#14 and \#49-\#52).
The Cable Feedthrough~2 and the Cable Pipe 2 were measured for the first time (\rnXenTNewCableFeedthrough \,and \rnXenTNewCablePipe, respectively).
Compared to Cable Pipe~1 (\#49 in~\cite{ref:xenon1t_radon}), Cable Pipe~2 has a factor $\sim\,$3 lower emanation rate.
This is consistent with the fact that total length of cable enclosed in Cable Pipe~2 is smaller with respect to Cable Pipe~1.
The integral rate of CRY is \rnXenTCryoIntegral \,mBq, excluding the emanation from the \mbox{XENONnT} TPC, which is discussed later.

The two GXe filter units of the \mbox{GXe-PUR} system were measured in their hot operating state (\rnXenTGXeGettera\, and \rnXenTGXeGetterb\, in~\cite{ref:xenon1t_radon}).
The emanation from the \mbox{Mag-Pump} 1 was taken from~\rnXenTGXeMaga\, in~\cite{ref:xenon1t_radon} as it was used 
already in the final science run phase of \mbox{XENON1T}.
The integral emanation value of the \mbox{GXe-PUR} system of \rnXenTGXeIntegral \,mBq, was obtained by adding the radon emanation from all individual parts.
The emanation of the second redundant \mbox{Mag-Pump}, serving as a backup system, is not considered.
\begin{figure*}[htbp]
\centering 
\includegraphics[width=0.7\textwidth,trim=0 60 0 60,clip]{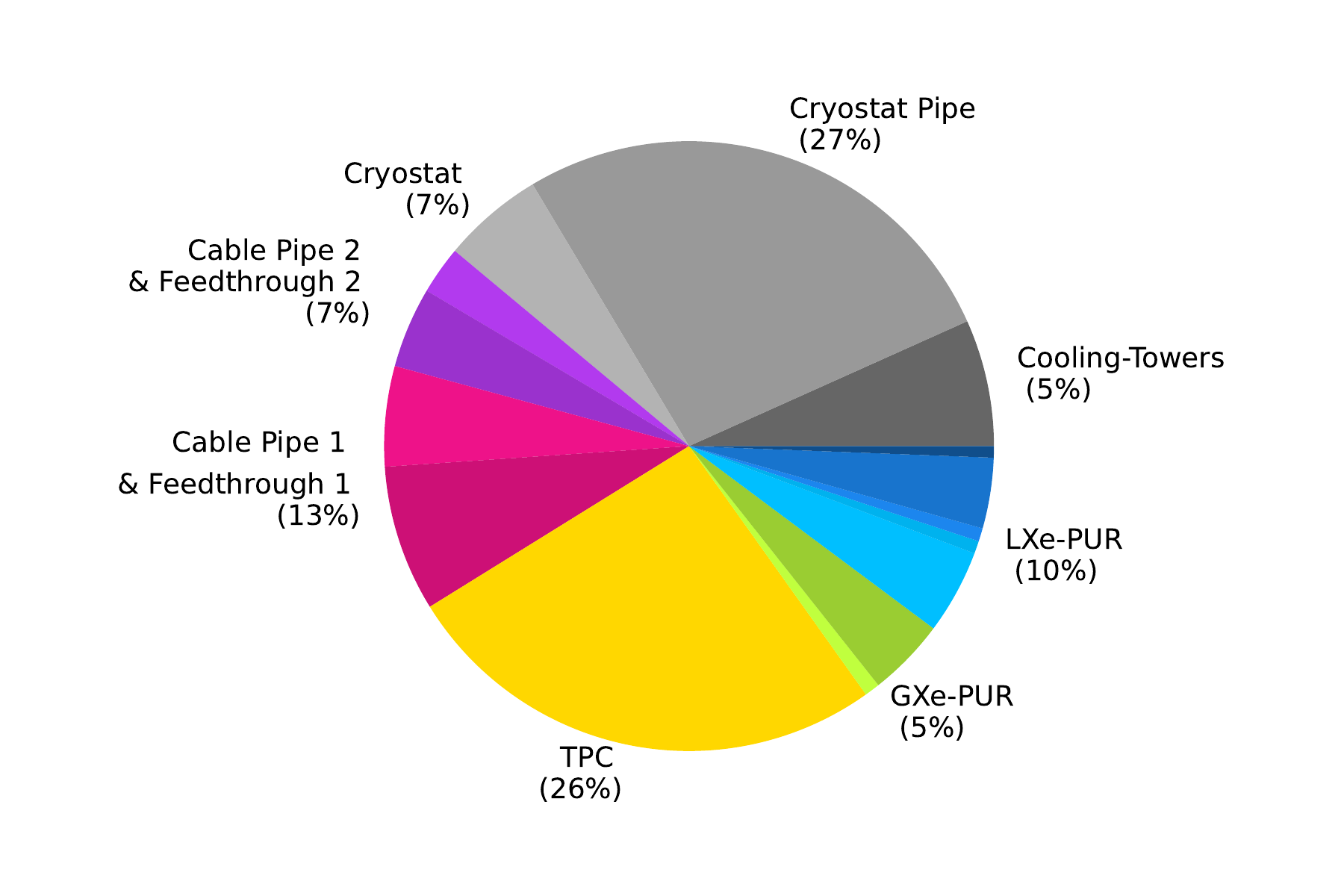}
\caption{The different subsystem contributions to the overall $^{222}$Rn emanation rate in \mbox{XENONnT}, adding up to \rnXenTTotal\,mBq. 
The colors correspond to the scheme used in 
Figure~\ref{fig:sketch_xenonnt}. 
Only central values from Table~\ref{tab:rn_XENONnT_samples} in this work and Figure~2 in~\cite{ref:xenon1t_radon} have been used.
The radon emanation from the \mbox{Rn-DST} system is not taken into account.}
\label{fig:pie_chart}
\end{figure*}

The \mbox{LXe-PUR} was split in six volumes which were measured separately. 
One volume (\rnXenTLXeVacumPipe) included the majority of vacuum insulated piping for the LXe and cryogenic valves. The second volume (\rnXenTLXePurityMonitor) contained the purity monitor which showed a similar emanation rate as the LXe filter 1 in the third volume (\rnXenTLXeFilterST). A 500W Xe-N$_{2}$ heat-exchanger was the main component of a forth volume (\rnXenTLXeHeatExchanger) which was measured to have a five times higher emanation rate than the previous three. 
The two last volumes contained one LXe pump each (\rnXenTLXePumpb). 
Both pumps gave similar results and were found to contribute to $\sim$44\% of the total radon emanation of \mbox{LXe-PUR}. 
Further investigations identified the pump's main-body made from SS as the main contributor.
Both pumps were dismounted and the main body of the pump was electropolished.
Earlier investigations showed a factor greater than three improvement may be achieved by electropolishing~\cite{ref:xenon1t_radon}.
A measurement of the pump's emanation rate after this treatment is not available. 
Therefore, the integrated result of \rnXenTLXeIntegral\,mBq for the \mbox{LXe-PUR} does not reflect the potential reduction and thus can only be seen as an upper limit.

An integral measurement of the \mbox{Rn-DST} system after assembly was not possible and the emanation rate of some single components remains undetermined.
The emanation of those items is estimated based on previous representative samples such as items \mbox{\#4-9} in~\cite{ref:xenon1t_radon} for SS and results from Table~1 in~\cite{ref:galex_prop_counters} for copper,  150\,(100)\,{\textmu}Bq/m$^{2}$ and 1.2\,(0.2)\,{\textmu}Bq/m$^{2}$, respectively. 
For the estimation of the distillation column's radon emanation rate (\rnXenTRADColumn), we consider the emanation from the eight SS packing-material pieces which fill the interior of the column. 
This packing material is the main contributor to the internal surface of the column. 
Their emanation rate is derived from the measurement of three packing-material pieces, given in Table~\ref{tab:rn_sulzer} (Pack 4/5/6 combined). The contribution of  other SS and copper surfaces in the distillation column to the total emanation rate is considered to be negligible.
Emanation from QDrive pump and GXe filter unit were also taken into account (\#20$_{a}$ in~\cite{ref:xenon1t_radon}, and \rnXeoneTRADgetter).
%
It should be noted that emanated radon atoms from the column, QDrive and the GXe filter 3 are expected to never reach the LXe target due to the distillation process.
Radon sources located downstream of the column, and thus after the radon removal process, 
are the four \mbox{Mag-Pumps} (\rnXenTRADMagcdef),  the related tubing (\rnXenTRADbufferin, \rnXenTRADbufferout\, and \rnXenTRADSecespol) and 
a heat exchanger (\rnXenTRADreboilerdown) where
radon-depleted xenon gas is liquified. 
The integral value of \mbox{Rn-DST} system is \rnXenTRnDSTIntegral\,mBq.

The most challenging measurement for \mbox{XENONnT} was the fully assembled TPC.
In the absence of a dedicated \mbox{gas-tight} vessel, the TPC could only be measured once enclosed in the cryostat together with the rest of the CRY system. 
Hence, the TPC emanation rate is an indirect measurement with respect to sample \textit{Integral CRY} in Table~\ref{tab:rn_XENONnT_samples}.
The measurement procedure was the same as for the rest of the samples described in this work (see Section~\ref{sec:222Rn_procedure});
however, due to the large PTFE surfaces and the expected outgassing, some extra precautions were taken in order to minimize the impact on the radon monitor's detector efficiency.
Firstly, before the measurement started, the entire CRY system and TPC were kept under vacuum pumping for several weeks to reduce the outgassing rate.
Secondly, a commercial N$_{2}$ gas purifier was installed at the extraction port such that the extracted N$_{2}$ carrier gas was further cleaned of impurities before reaching the radon trap.
The averaged TPC emanation rate obtained from three separate, consecutive radon extractions is \rnXenTtpcIntegralStat\,mBq.

For the TPC measurement, the carrier gas was extracted through several ports located at the bell, Cable Feedthrough~2, Cryostat Pipe and cooling towers.
The amount of gas extracted via each individual port was varied between the three measurements.
An active mixing of the carrier gas prior to extraction was not possible.
Thus a \mbox{non-homogeneous} radon concentration potentially influenced the TPC result due to the applied scaling. 
This systematic uncertainty was estimated in a numeric simulation that accounts for the known emanation rates of detector subsystems and other details about the extraction procedure (i.e., the locations of the extraction ports and their individual gas flows). 
The simulation was developed to probe different scenarios of the dynamics of the carrier gas before and during the extraction, ranging from a laminar flow to a turbulent mixing of the carrier gas in all detector \mbox{subvolumes}.
The final TPC emanation rate, based on this systematic uncertainty study, was found to be \rnXenTtpcIntegral\,mBq (\rnXenTtpc\,in Table~\ref{tab:rn_XENONnT_samples}).
To put this value into context, the emanation from the 494 PMTs is expected to be of $\sim\,$2.2\,mBq (from \#37 and \#38 in~\cite{ref:xenon1t_radon}), and the emanation from cables running inside the inner cryostat add additional $\sim\,$3\,mBq (as discussed already in Section~\ref{sec:222Rn_items}).
The emanation from other TPC materials, such as PTFE or copper, was not measured prior to assembly.

Taking into account the presented radon emanation measurements for the different xenon handling systems, the total $^{222}$Rn\, emanation\, rate\, of\, \mbox{XENONnT} is estimated to be \rnXenTTotal\,mBq. 
It should be noted that this number is estimated from measurements performed at ambient temperature where the radon emanation rate might be increased with respect to the emanation at the detector's operating temperature.  For XENON1T, however, this effect was not observed as the measured radon concentration in the operating TPC was about 30\% higher with respect to the expectation from radon emanation measurement~\cite{ref:xenon1t_radon}.
A further $^{222}$Rn reduction due to the operation of the \mbox{Rn-DST} system is not considered here.
Figure~\ref{fig:pie_chart} shows how the sources contribute to the overall radon budget. 
Assuming a homogeneous distribution of radon in the entire xenon inventory, the total radon emanation rate translates to an activity concentration of \rnXenTConcentration\,{\textmu}Bq/kg.
The radon concentration in LXe is expected to be reduced by the operation of \mbox{Rn-DST}, 
bringing the \mbox{XENONnT} target value of 1\,{\textmu}Bq/kg within reach. 

\section{Surface treatment}
\label{sec:clean}
A thorough cleaning of materials serves multiple purposes. It removes small particulates that, if released inside the detector, might compromise the detector's operation, in particular the HV stability of the electrodes. Furthermore, the cleaning process removes grease and lubricants that may be left over from the manufacturing process. 
An adequate chemical treatment also leads to controlled passivation and surface conditioning of delicate metallic surfaces, such as the detector's fine electrode wires \cite{ref:electron_emission}. Finally, a dedicated cleaning procedure can help to remove radioactive isotopes that have been accumulated on material surfaces during production, storage and handling. 
This section summarizes the cleanliness efforts in preparation for the \mbox{XENONnT} detector, including cleaning procedures, infrastructure, and the measures taken to avoid re-contamination during detector assembly.

\subsection{Cleanroom infrastructure}
\label{sec:cr_infrastructure}
For material cleaning, storage and detector assembly, two cleanrooms (CRs) at the LNGS laboratory were utilized.
The above-ground cleanroom (AG-CR) was located inside an assembly hall and had a footprint of $(9\times 5)$\,m$^2$. The ambient air was cleaned using HEPA filters and then flushed with laminar flow through the CR. Periodic particle counter measurements demonstrated ISO 6 classification. All of the large-scale detector materials, including the TPC, were cleaned in the AG-CR. In order to maintain cleanliness, all materials entered the cleanroom via an anteroom where pre-cleaning happened (see Section~\ref{sec:cr_procedures}). Large-scale items were brought directly into the AG-CR through a gate bypassing the anteroom. Thanks to the adjusted air flow and movable curtains, which were mounted before opening the gate, a temporary anteroom could be established for the pre-cleaning of these large items. 

\begin{figure}[htbp]
\centering 
\includegraphics[width=0.4\textwidth]{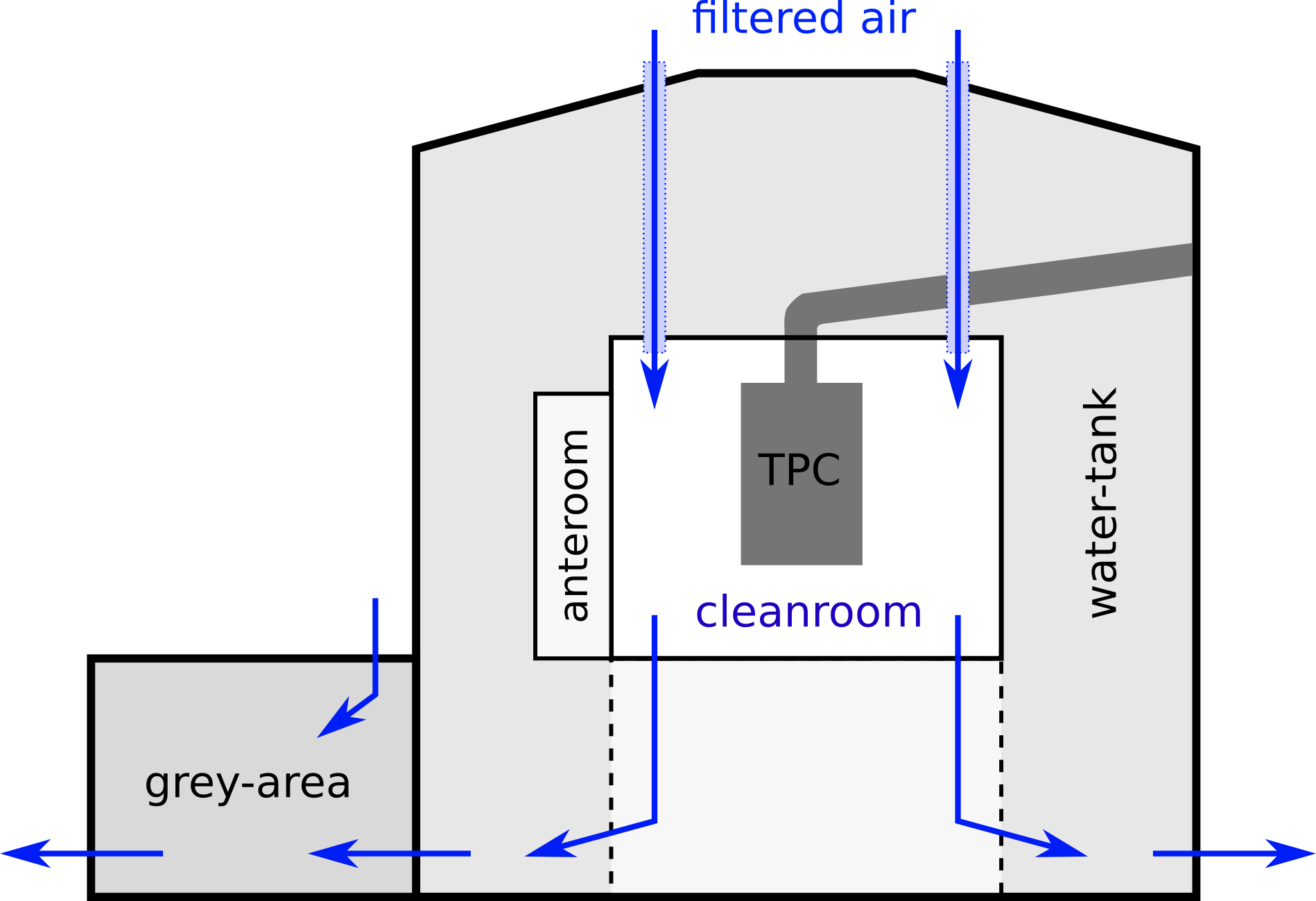}
\caption{The UG-CR (ISO 6 class) is built around the cryostat in the center of the water tank. Filtered air is flushed first into the cleanroom in the center and is pushed thereafter through the water tank into the grey area, providing a sequence of decreasing cleanliness levels toward ambient air. }
\label{fig:ug_cr}
\end{figure}

After assembly in the AG-CR, the TPC was protected against contamination (see Section~\ref{sec:mat_storage}) and transported to the experimental site underground for installation. In order to guarantee a clean installation, a dedicated underground cleanroom infrastructure (UG-CR) of different cleanliness levels was built inside \mbox{XENONnT}'s water tank, as shown in Figure~\ref{fig:ug_cr}. The UG-CR was accessible through a so-called grey area that was temporarily built in front of the water tank during the construction phase of \mbox{XENONnT}. Constantly flushed with filtered air, it was used to clean materials and equipment before entering the CR. From the grey area, one could access the water tank, which as a whole was provided with filtered air at a slightly higher pressure than in the grey area. The water tank was treated as a clean environment. 
Materials and equipment foreseen for the CR were cleaned or unpacked here.
The actual CR had a footprint of $(5\times 5)$\,m$^2$ and was located at the center of the water tank. It enclosed the \mbox{XENONnT} cryostat after the cryostat's assembly. Filtered air was flushed directly into the CR, guaranteeing a certified ISO 6 classification. From there, the clean air was guided through the water tank and the grey area following the decreasing cleanliness levels of those areas. 
When needed, the area right below the CR could be separated from the rest of the water tank by movable plastic curtains (shown with dashed lines in Figure~\ref{fig:ug_cr}). The floor of the CR was then opened to lift the assembled TPC or the inner cryostat vessel up into the CR. The filtered air was flushed with high flow rate from the CR through the separated area underneath to mantain the air quality throughout this temporarily enlarged volume. 

\subsection{Cleaning procedures}
\label{sec:cr_procedures}
All final cleaning happened in a clean environment, mostly in the AG-CR. Deionized water with an average electrical conductivity of $\sim \,$0.08\,{\textmu}S/cm was supplied by a dedicated water plant with a flow up to $1.3\,$m$^3$/h. For the cleaning of large-scale items, the AG-CR was equipped with two containers made of high-density polyethylene (HDPE) with a $(1.9\,\times 1.6\,\times 0.4)\,$m$^3$ volume and a slightly larger container of the same type. The latter was used as a custom-made US-bath and housed four US transducers each with a $(0.5 \times 0.7)\,$m$^2$ surface area and a maximum power of $2\,$kW at a frequency of $40$\,kHz. Two heaters of $20\,$kW in total were available to bring the baths to the required temperatures. 
A custom-made crane was built to handle heavy items during cleaning.
Prior to bringing detector materials, tools, and containers into any CR, they were pre-cleaned with ethanol-soaked wipes or neutral soap. Further degreasing followed as a first cleaning step inside the CR. Therefore, detergents were selected depending on the materials to clean but also with respect to their internal radioactivity (see Table~\ref{tab:screening}). Alkaline detergents such as HARO Clean~188\footnote{https://harosol.com/en/} (item 34 in Table
~\ref{tab:screening}) showed high $^{40}$K activities of up to several kBq/kg and were not applied to soft or porous materials such as PTFE, which might absorb small amount of the solution including the radioactive impurities. For stainless steel, gamma spectroscopy measurements did not indicate an increase of the material's activity after the usage of such solutions. Attention was also paid to the detergents' $^{226}$Ra contamination. Being the progenitor of $^{222}$Rn, radium plating out on the material surfaces during the cleaning process could significantly increase the radon emanation rate. For a sample of P3-Almeco 36\footnote{www.henkel-adhesives.com}, a $^{226}$Ra contamination of $98(27)\,$mBq/kg was found (see item 36 in Table~\ref{tab:screening}), the highest of all screened detergents.
Indications for the plate-out of radium were observed during the cleaning process of samples of the SS packing material used in the cryogenic radon distillation column (see Table~\ref{tab:rn_sulzer}).

\begin{table*}[t]
\centering
\renewcommand{\arraystretch}{1.0}
\smallskip
\begin{tabular}{l | l c}
\hline
\textbf{Sample} & \textbf{Procedure} & \textbf{Emanation rate [mBq/piece]} \\
\hline
Pack 1 & no cleaning & $0.13(4)$ \\
 & Almeco treatment & $0.80(5)$ \\
 \hline
Pack 2 & Almeco treatment & 1.30(6)\\
       & additional DI water rinsing & 0.92(7)\\
       & repeat additional DI water rinsing & 0.86(10)\\
\hline
Pack 3 & Almeco treatment & 2.58(15)\\
       & repeat Almeco treatment & 3.26(18)\\
\hline
Pack 4/5/6 combined & Acetone treatment & 0.21(3)\\
\hline
\end{tabular}
\caption{\label{tab:rn_sulzer} Radon emanation results of distillation packing material before and after selected cleaning procedures.}
\end{table*}
After a 10-minute bath in 5\% P3-Almeco 36 solution at $60\,^\circ$C, one sample's radon emanation rate increased by a factor of six to $0.80(5)\,$mBq (Pack~1 in Table~\ref{tab:rn_sulzer}). Other samples (Pack~2 and Pack~3) also showed an emanation rate significantly higher than that of the untreated samples. In the case of Pack~3, the repetition of the P3-Almeco 36 cleaning procedure further increased its emanation rate. Attempts to use DI water to flush out or dilute potential detergent residuals from the largely inaccessible, heavily folded surface yielded only a small reduction (Pack~2). The packing material was cleaned in acetone instead, for which no $^{226}$Ra contamination was detected after treatment. For the cleaning of other large volume materials, mostly Elma clean 65\footnote{www.elma-ultrasonic.com} was used (see item 38 in Table~\ref{tab:screening}), the detergent with the lowest measured $^{226}$Ra contamination, as well as a mixed solution from the detergents HARO Clean~188 and HARO Clean~106 (see items 34 and 35 in Table
~\ref{tab:screening}).

After the first detergent treatment, most detector parts were cleaned in an acidic solution adjusted to their material composition, as shown below. Depending on the material, the purpose of this cleaning step is either to remove a thin surface layer from the material, and thus its contamination, or to dissolve impurities in the solution. The purity level of the chemicals was $>$99\% (pro analysis).
A gamma spectroscopy measurement of a nitric acid sample was used to set upper limits on its radiopurity, most importantly for $^{226}$Ra (see item 40 in Table~\ref{tab:screening}). The detailed procedures for different types of materials are specified below.

\subsubsection*{Copper}
Copper cleaning was developed based on a procedure described in \cite{ref:copper_cleaning} for the removal of surface contamination while retaining surface details such as threads or boreholes.
\begin{itemize}
    \item Elma clean 65 neutral soap (5\%), $15\,$min at $35-40\,^\circ$C in US-bath
    \item thorough DI water rinsing
    \item H$_2$SO$_4$ (1\%) + H$_2$O$_2$ (3\%) solution, $5\,$min at room temperature
    \item immerse in DI water bath
    \item citric acid (5\%), $5\,$min at room temperature
    \item thorough DI water rinsing
    \item cleanroom wipes and N$_2$ blowing for drying
\end{itemize}
A re-deposition of dissolved copper in the sulfuric acid solution could be prevented by moving the copper pieces throughout the washing. The drying process needed to happen immediately after the final rinsing to prevent a rapid oxidation of the copper. For large items, an increased temperature of the last rinsing bath ($\sim 35^{\circ}$C) supported this drying process with its implied higher evaporation rate.

\subsubsection*{PTFE and other plastics}
PTFE is known to attract positively charged radon daughters and thus promote their plate-out on its surface \cite{ref:ptfe_plateout}. In order to increase the reflectivity of vacuum ultraviolet light, the PTFE surfaces that face the inner LXe volume have been shaved with a diamond-tipped tool. During this process up to 1.5\,mm of surface layer was removed. Thus, impurities located on the material surface or just below were removed, making the shaving process an important cleaning step. This effect was studied using the XIA UltraLo spectrometer\footnote{www.xia.com/ultralo.html}, located in the underground laboratory at Kamioka \cite{ref:xia} which measured the surface activity of $^{210}$Po on PTFE samples. For an unshaved reflector sample, an activity of $126(8)\,$mBq/m$^2$ was measured, while a reduction to $20(3)$\,mBq/m$^2$ was observed on the PTFE-surface treated with the diamond-tipped tool. Only the surfaces of PTFE reflector panels facing the TPC underwent the shaving process.\\

All PTFE parts, but also other plastics such as Kapton and PEEK were chemically cleaned according to the procedure outlined below. The capability of different cleaning procedures to remove radon daughters from PTFE surfaces was studied in~\cite{ref:rn_daughters}. It was shown that the contamination of $^{210}$Po and $^{210}$Pb could be reduced by a factor two at most, almost independent of the chemicals used. This indicates that only radon daughters that have not been implanted into the bulk PTFE can be removed. A better reduction factor of $\sim \,$30 has been achieved for the removal of $^{212}$Pb from PTFE surfaces~\cite{ref:rn_daughters}. Thus, the following nitric acid based procedure was used:
\begin{itemize}
    \item Elma clean 65 neutral soap (5\%), $15\,$min at $35-40\,^\circ$C in US-bath (no US used for diamond-shaved parts)
    \item thorough DI water rinsing
    \item HNO$_3$ (5\%) solution, $2\,$h at room temperature including 15\,min in US-bath (no US used for diamond-shaved parts)
    \item immerse in DI water bath up to 1\,h to dissolve acid residuals
    \item thorough DI water rinsing
    \item N$_2$ blowing for drying
\end{itemize}
For the diamond-shaved surfaces of the PTFE reflector panels the US-bath was not used to avoid the risk of damaging the treated surfaces. 
A dedicated storage in nitrogen-flushed boxes helped to mitigate the re-contamination of the PTFE with radon daughters (see Section~\ref{sec:mat_storage}) and removed residual humidity.

\subsubsection*{Electrodes and other stainless steel items}
The five electrodes of \mbox{XENONnT} are made from SS wires of diameters $216$\,{\textmu}m and $304$\,{\textmu}m, fixed to a corresponding SS electrode frame. Before assembly of the electrodes, the wires were mounted on a holding structure designed for the cleaning process. The electrode frames themselves were cleaned separately, using the same procedure as for the wires. As a detergent, a mixture of the alkaline HARO Clean~188 and the cleaning amplifier HARO Clean~106 was used. A positive effect of nitric acid (35\%) on the emission of single electrons from electrodes was documented in \cite{ref:electron_emission}. For safety reasons, however,  7\% citric acid solution was used instead, an alternative approach that is common in industry for SS passivation when nitric acid cannot be utilized. The detailed procedure is:
\begin{itemize}
    \item HARO Clean 188 (5\%) + HARO Clean 106 (0.01\%) solution, $10\,$min at $45-50\,^\circ$C in US-bath
    \item thorough DI water rinsing
    \item Citric acid (7\%) solution, $1\,$h at $(45-50)\,^\circ$C including 10\,min in US-bath
    \item thorough DI water rinsing
    \item storage in N$_2$ flushed boxes for drying
\end{itemize}
After assembly, the wired electrodes were cleaned again in Elma clean 65 (5\%) solution for 15\,min at room temperature, without US.
Other small SS parts were treated only with the HARO Clean~188/106 detergent mixture (first step in the procedure above). Large SS items were electropolished by the manufacturers beforehand in order to remove surface contamination \cite{ref:rn_electropolish1,ref:rn_electropolish2}. The inner cryostat vessel was cleaned with ethanol-soaked wipes after its electropolishing.

\subsubsection*{PMTs and cables}
During the decommissioning of the \mbox{XENON1T} detector, its PMTs were dismounted and sealed in a CR environment for later use in \mbox{XENONnT}. All PMTs, including the newly purchased ones, were cleaned inside the AG-CR before the detector assembly. The procedure included:
\begin{itemize}
    \item N$_2$ blowing to remove small particulates
    \item soft wiping of insensitive parts using ethanol
    \item immersion of the PMT in an ethanol bath
\end{itemize}
HV and signal cables have been soldered to the PMT bases beforehand. Therefore, cabels and bases needed to be cleaned together. For better handling, the bases were mounted on acrylic structures to keep them in place throughout the cleaning process which included the following steps:
\begin{itemize}
    \item thorough DI water rinsing to remove dust particles before entering the AG-CR
    \item immerse only the cables in Elma clean 65 soap (5\%), 15\,min at 30$\,^{\circ}$C in US-bath, the bases stayed outside the bath
    \item immerse also bases in Elma clean 65 soap (5\%), 15\,min at 30$\,^{\circ}$C without US
    \item thorough DI water rinsing
    \item N$_2$ blowing for drying
    \item immerse cables and bases in ethanol bath to accelerate the drying process
\end{itemize}
After drying, cables and bases were wrapped and stored inside the CR until assembly.

Once installed in the TPC, the cables, which are directly connected to the PMT bases, reach only from the PMT array to the opening of the two cable pipes at the inner dome of the cryostat (see Figure~\ref{fig:sketch_xenonnt}). There they were connected to cables which had been pre-installed inside the two cable pipes. The cables in the cable pipe~1 were reused from \mbox{XENON1T} and kept under nitrogen atmosphere during the detector upgrade so that no further cleaning was necessary. The cables in the new Cable Pipe~2 were thoroughly wiped with ethanol inside a CR before being installed in the pipe.

\subsection{Plate-out of radioactive impurities and material storage}
\label{sec:mat_storage}

Detector materials must be protected and stored properly to maintain cleanliness. Besides the plate-out from generic dust a particular emphasis was placed on mitigating contamination from radium and radon daughters on PTFE surfaces. In order to estimate the plate-out rate of long-lived radon daughters, PTFE plates of 20\,cm$^2$ surface area were placed at various locations in the CRs and at the \mbox{XENONnT} experimental site. By means of alpha spectroscopy~\cite{ref:rn_daughters} an upper limit for the increase of the surface activity of $^{210}$Po due to plate-out of $<$19\,mBq/d/m$^2$ (95\% C.L.) was determined.
To investigate potential radium plate-out e.g. from dust in ambient air, two PTFE samples, each of $3\,$m$^2$ surface area, were exposed to ambient air at two different locations . After 105 days of exposure, the increase of the $^{222}$Rn emanation rate due to the radium plate-out was measured by employing the radon assay technique described in Section~\ref{sec:radon}. For the sample placed at the \mbox{XENONnT} experimental site, an increase of the $^{222}$Rn emanation rate of $0.28(15)$\,{\textmu}Bq/d/m$^2$ was observed. The averaged radon concentration in the ambient air during the exposure time was $36\,$Bq/m$^3$.
\begin{figure}[htbp]
\centering 
\includegraphics[width=0.35\textwidth]{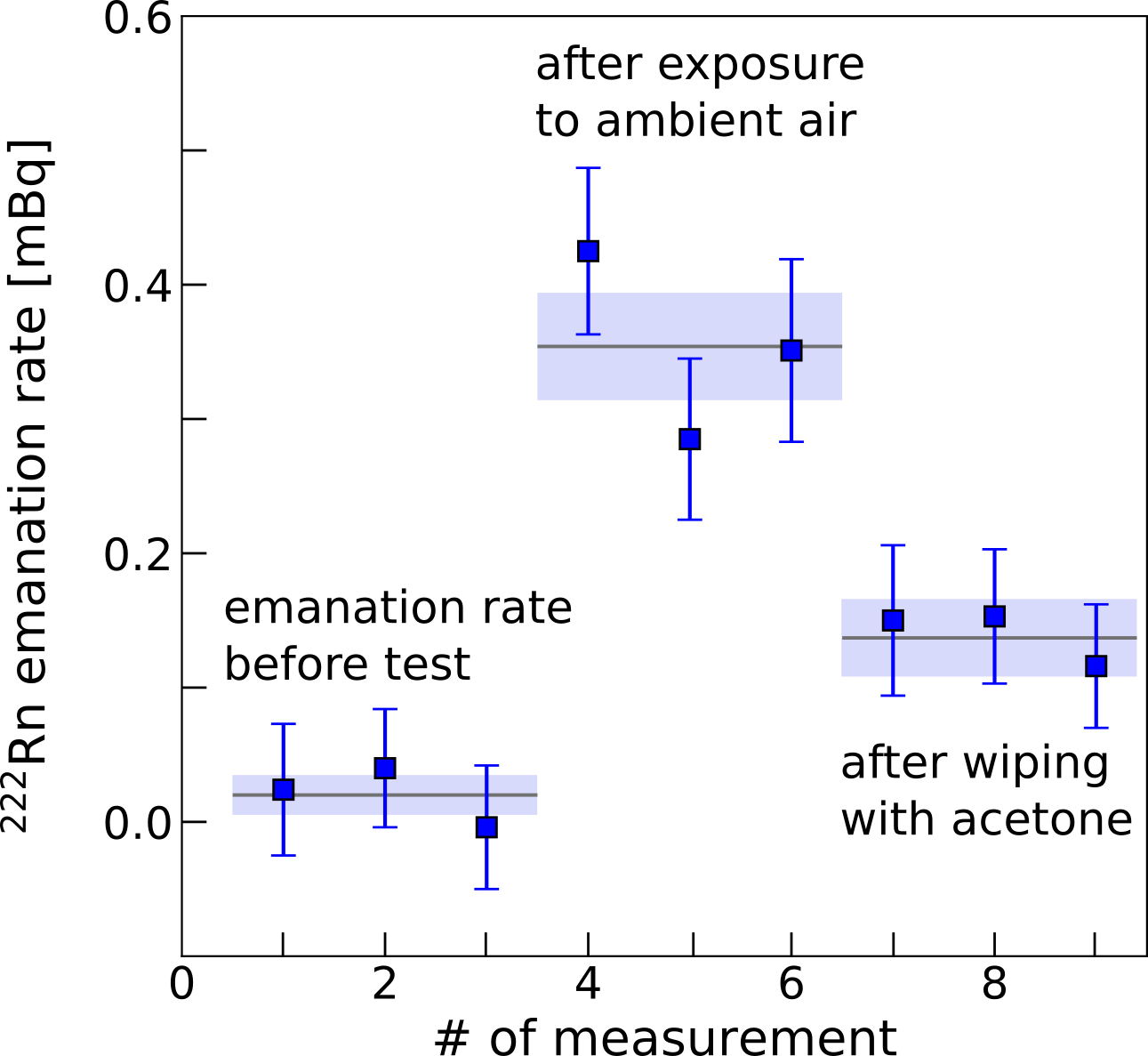}
\caption{Calculated radon emanation rate after background subtraction of a 3\,m$^2$ large PTFE foil before and after a 105\,d exposure to ambient air with a radon concentration of $130\,$Bq/m$^3$. The initial emanation rate was not reached again after wiping the sample with acetone-soaked cleanroom wipes.}
\label{fig:ptfe_curtain}
\end{figure}
The second sample was placed in a room with a radon concentration in ambient air of about $130\,$Bq/m$^3$, a factor 3.6 higher with respect to the \mbox{XENONnT} experimental site. The determined $^{222}$Rn emanation rate increase was about the same factor higher and was determined to be $1.05(14)$\,{\textmu}Bq/d/m$^2$. The aerosol concentration in ambient air, which might also impact the plate-out rate, was about $10^6$\,particles/m$^3$ (diameter $>0.5$\,{\textmu}m) for the second sample. After measuring the radon emanation rate the second sample was cleaned with acetone-soaked cleanroom wipes, which reduced the emanation rate by $\sim 60$\%, but still well above the initial value (see Figure~\ref{fig:ptfe_curtain}). The process of $^{226}$Ra plate-out needs further investigation. Dust from ambient air is thought to be the main source but comprehensive studies on the plate-out mechanisms do not yet exist.


For plate-out protection, PTFE parts, PMTs and electrodes, were stored in nitrogen flushed boxes or were sealed in bags made of mylar foil. For the latter, their radon tightness could be proven by sealing a sample of $>$30\,mBq radon emanation rate in such a mylar foil and placing the bag in a vessel filled with helium carrier gas. After one week, the carrier gas was analysed as it was done for ordinary radon emanation measurements. No significant amount of radon was detected. Mylar bags were also used during the transport of the TPC from the assembly site in the AG-CR to its final destination in the UG-CR, as well as during the installation phase at the experiment site.

\section{Summary and conclusions}
\label{sec:results}

To achieve the target background level in \mbox{XENONnT}, an extensive screening campaign was performed to select construction material with low intrinsic radioactivity. Their gamma emission was measured with high-purity germanium detectors with sensitivities down to $\sim10$\,{\textmu}Bq/kg. \mbox{ICP-MS} was employed for complementary measurements of the $^{238}$U and $^{232}$Th concentrations in the materials. 
The isotopic activities measured in the radioassay program informed the simulated sensitivity estimates for \mbox{XENONnT} for low-energy nuclear recoils published in~\cite{ref:xenonnt_mc}.
Relative to \mbox{XENON1T}, the selection of materials, as described here, combined with the increased efficacy of fiducialization lead to a reduction ($\sim$17\%) in the ER background contribution from detector components, contributing $25(3)$\,events per tonne-year in \mbox{XENONnT}.
This contribution is subdominant to $^{222}$Rn and solar neutrinos, contributing with $55(6)$ and  $34(1)$ events respectively (assuming 1\,{\textmu}Bq/kg for $^{222}$Rn). 
Similarly, the NR background from detector components, $0.32(0.16)$\,events per tonne-year, was cut in half compared to \mbox{XENON1T}. The most significant contributors are the SS cryostat (36\%), PMTs (33\%), and PTFE components (26\%). The neutron veto, which is expected to tag radiogenic events with $\sim$87\% efficiency, should further mitigate the materials-induced background to $0.04\,(0.02)$\,events per tonne-year. 
Consequently, radiogenic neutrons would no longer be the dominant source of NR events. The solar $^8$B CEvNS events instead should constitute the largest population of NR background events for dark matter searches, reflecting notable improvements in sensitivity, while also providing their own novel channel of investigation.\\ 

As the radioactive noble gas $^{222}$Rn is expected to be the dominant ER background source in \mbox{XENONnT}, the materials have been screened also for their radon emanation rate.
Furthermore, the $^{222}$Rn emanation rate of entire detector subsystems was measured during the assembly of \mbox{XENONnT}. 
Based on these measurements, the locations of the main radon sources could be identified. This knowledge will be used to optimize the efficiency of a novel radon removal system based on cryogenic distillation.
The largest $^{222}$Rn source in the \mbox{XENONnT} experiment is the emanation from its cryogenic system. 
The TPC, with a total activity of \rnXenTtpcIntegral\, mBq, is the second largest source of radon.
The total emanation in \mbox{XENONnT} was estimated to be \rnXenTTotal\,mBq. 
Assuming a homogeneous distribution within the \rnXenTInventory\,tonnes of xenon in the detector, a final radon activity concentration of \rnXenTConcentration\,{\textmu}Bq/kg in the LXe target is expected, 
a factor of three lower than in \mbox{XENON1T}.
\mbox{XENONnT}'s novel radon distillation system will further reduce the radon concentration in LXe, 
allowing us to achieve the target activity of 1 {\textmu}Bq/kg. 
Imminent \mbox{XENONnT} TPC data will validate this post-distillation radon concentration~projection.\\

Special emphasis was placed on the cleaning and proper storage of all detector materials during the assembly of \mbox{XENONnT}. Dedicated cleaning procedures were defined for different materials in order to remove dust and lubricants introduced during the material production process. The cleaning agents were selected 
according to results obtained from gamma-ray screening. For some materials, an increased radon emanation rate was detected after their degreasing treatment, which was associated with a relatively high radium concentration in the cleaning agent. Surface treatments are also important to mitigate background from long-lived radon daughters. 
For PTFE surfaces, the most critical component in this regard for \mbox{XENONnT}, a reduction up to a factor of six was established for the procedure described above. 
In order to avoid re-contamination of detector parts, 
a dedicated cleanroom infrastructure was built for material cleaning, storage, and detector assembly.

\section*{Acknowledgements}
We gratefully acknowledge support from the National Science Foundation, Swiss National Science Foundation, German Ministry for Education and Research, Max Planck Gesellschaft, Deutsche Forschungsgemeinschaft, Helmholtz Association, Dutch Research Council (NWO), Weizmann Institute of Science, Israeli Science Foundation, Fundacao para a Ciencia e a Tecnologia, R\'egion des Pays de la Loire, Knut and Alice Wallenberg Foundation, Kavli Foundation, JSPS Kakenhi in Japan and Istituto Nazionale di Fisica Nucleare. This project has received funding/support from the European Union’s Horizon 2020 research and innovation programme under the Marie Sk{\l}odowska-Curie grant agreement No 860881-HIDDeN. Data processing is performed using infrastructures from the Open Science Grid, the European Grid Initiative and the Dutch national e-infrastructure with the support of SURF Cooperative. We thankfully acknowledge the work of the Super-Kamiokande gadolinium group on developing a source of radio-pure gadolinium sulfate with the Nippon Yttrium Co. LTD. (NYC) in Japan and giving us access to their product. We further acknowledge productive collaboration in developing suitable gadolinium sulfate for XENONnT by Treibacher Industrie AG in Austria. We are grateful to Laboratori Nazionali del Gran Sasso for hosting and supporting the XENON project.

\input{biblio}

\onecolumn

\thispagestyle{plain}
\pagenumbering{gobble}

\begin{landscape}
\begin{table}[h]
\scriptsize
\vspace{3mm}
\begin{tabular}{l l l l c c c c c c c c c c c c}
\textbf{Sample} & \textbf{Component}& \textbf{Manufacturer}& \textbf{Facility}& \textbf{Mass [kg]}& \textbf{Livetime [d]}& \textbf{Units}& \textbf{$^{238}$U}& \textbf{$^{235}$U} & \textbf{$^{226}$Ra}& \textbf{$^{228}$Ra ($^{232}$Th)}& \textbf{$^{228}$Th}& \textbf{$^{40}$K}& \textbf{$^{60}$Co}& \textbf{$^{137}$Cs}\\
\hline 
\noalign{\vskip 2mm}
\multicolumn{3}{l}{\textbf{Stainless Steel (304)}} &&&&&&&&&&&&&\\ 
\hline
0 & Bell/Vessel & Nironit & GeMPI & 7.8 & 11.7 & mBq/kg & $13(7)$ & $0.7(3)$ & $0.3(1)$ & $0.6(2)$ & $0.5(1)$ & $1.6(6)$ & $2.4(2)$ & $<0.2$ \\ 
0 & Bell/Vessel & Nironit & ICP-MS & $-$ & $-$ & mBq/kg & $3.7(6)$ & $-$ & $-$ & $0.10(8)$ & $-$ & $-$ & $-$ & $-$ \\ 
1 & Bell/Vessel & Nironit & GeMPI & 7.8 & 57.1 & mBq/kg & $4(2)$ & $0.2(1)^{*}$ & $1.3(1)$ & $0.9(1)$ & $0.57(6)$ & $1.4(2)$ & $0.61(5)$ & $0.03(2)$ \\ 
1 & Bell/Vessel & Nironit & ICP-MS & $-$ & $-$ & mBq/kg & $8.6(4)$ & $-$ & $-$ & $<8.1$ & $-$ & $-$ & $-$ & $-$ \\ 
2 & Bell/Vessel/Electrodes & Nironit & GeMPI & 8.4 & 27.5 & mBq/kg & $<11$ & $<0.6$ & $0.6(1)$ & $0.4(1)$ & $0.4(1)$ & $<2.4$ & $0.4(1)$ & $<0.2$ \\ 
2 & Bell/Vessel/Electrodes & Nironit & ICP-MS & $-$ & $-$ & mBq/kg & $2.5(3)$ & $-$ & $-$ & $0.4(2)$ & $-$ & $-$ & $-$ & $-$ \\ 
3 & Welding Rods (Vessel) & Nironit & GeMPI & 2.6 & 30.6 & mBq/kg & $<5.7$ & $<0.3^{*}$ & $3.1(3)$ & $2.9(4)$ & $11.4(7)$ & $7(1)$ & $1.6(2)$ & $<0.3$ \\ 
\multicolumn{3}{l}{\textbf{Oxygen-Free High-Conductivity Copper}} &&&&&&&&&&&&&\\
\hline
4 & Field Shaping Rings & Luvata & Gator & 71.7 & 32.5 & mBq/kg & $<0.33$ & $<0.02$ & $<0.18$ & $<0.22$ & $0.18(5)$ & $0.45(14)$ & $0.03(1)$ & $<0.05$ \\ 
4 & Field Shaping Rings & Luvata & ICP-MS & $-$ & $-$ & mBq/kg & $0.03(1)$ & $-$ & $-$ & 0.010(4) & $-$ & $-$ & $-$ & $-$ \\ 
5 & Guard Rings & Niemet & GeMPI & 56.5 & 42.1 & mBq/kg & $<1.6$ & $<0.14$ & $0.13(3)$ & $<0.06$ & $<0.04$ & $0.6(2)$ & $0.05(1)$ & $<0.03$ \\ 
6 & Wires & - & GeMSE & 12 & - & mBq/kg & $<2.3$ & $-$ & $<0.1$ & $<0.06$ & $<0.04$ & $0.55(2)$ & $0.43(3)$ & $<0.04$ \\ 
7 & Array Support Plate & Niemet & GeMSE & 93.4 & 35.6 & mBq/kg & $<1.06$ & $-$ & $<0.21$ & $<0.08$ & $<0.01$ & $<0.42$ & $0.08(1)$ & $<0.011$ \\ 
7 & Array Support Plate & Niemet & ICP-MS & $-$ & $-$ & mBq/kg & $0.0014(4)$ & $-$ & $-$ & 0.004(1) & $-$ & $-$ & $-$ & $-$ \\ 
8 & Array Support Pillar & Luvata & GeMPI & 57.3 & 26.2 & mBq/kg & $<2.7$ & $<0.23$ & $<0.06$ & $<0.08$ & $<0.04$ & $<0.27$ & $0.10(2)$ & $<0.05$ \\ 
\multicolumn{3}{l}{\textbf{Plastics}} &&&&&&&&&&&&&\\
\hline
9 & PTFE Reflectors & Amsler \& Frey & GeMPI & 15.4 & 25.0 & mBq/kg & $<2.4$ & $<0.08$ & $<0.03$ & $0.11(4)$ & $<0.09$ & $8(1)$ & $-$ & $<0.07$ \\ 
9 & PTFE Reflectors & Amsler \& Frey & ICP-MS & $-$ & $-$ & mBq/kg & $<0.06$ & $-$ & $-$ & 0.05(2) & $-$ & $-$ & $-$ & $-$ \\ 
10 & PTFE Reflectors & Amsler \& Frey & GeMPI & 25.0 & 19.7 & mBq/kg & $<1.0$ & $<0.07$ & $0.15(3)$ & $<0.1$ & $<0.08$ & $0.08(3)$ & $-$ & $<0.05$ \\ 
10 & PTFE Reflectors & Amsler \& Frey & ICP-MS & $-$ & $-$ & mBq/kg & $0.15(7)$ & $-$ & $-$ & 0.03(2) & $-$ & $-$ & $-$ & $-$ \\ 
11 & PTFE Pillars & Amsler \& Frey & GeMPI & 15.1 & 45.0 & mBq/kg & $<0.8$ & $<0.05$ & $0.04(1)$ & $<0.06$ & $<0.04$ & $<0.42$ & $-$ & $<0.01$ \\ 
11 & PTFE Pillars & Amsler \& Frey & ICP-MS & $-$ & $-$ & mBq/kg & $0.26(9)$ & $-$ & $-$ & $0.10(2)$ & $-$ & $-$ & $-$ & $-$ \\ 
12 & PTFE PMT Holders & Amsler \& Frey & GeMSE & 18.2 & 19.9 & mBq/kg & $<1.9$ & $-$ & $<0.1$ & $<0.08$ & $<0.04$ & $<1.0$ & $<0.05$ & $<0.03$ \\ 
13 & PTFE PMT Holders & Amsler \& Frey & ICP-MS & $-$ & $-$ & mBq/kg & $<0.1$ & $-$ & $-$ & $<0.04$ & $-$ & $-$ & $-$ & $-$ \\ 
14 & Torlon Reflectors & Drake Plastics & ICP-MS & $-$ & $-$ & mBq/kg & $1.8(5)$ & $-$ & $-$ & $0.2(1)$ & $-$ & $-$ & $-$ & $-$ \\ 
15 & Torlon Reflectors & Drake Plastics & ICP-MS & $-$ & $-$ & mBq/kg & $2.2(6)$ & $-$ & $-$ & $0.4(1)$ & $-$ & $-$ & $-$ & $-$ \\ 
16 & PEEK Array Spacers & Spalinger & ICP-MS & $-$ & $-$ & mBq/kg & $0.4(1)$ & $-$ & $-$ & $0.12(3)$ & $-$ & $-$ & $-$ & $-$ \\ 
17 & PEEK Screws & Solidspot & GeMPI & $0.27$ & $23.1$ & mBq/kg & $<20$ & $<1.4$ & $10(1)$ & $7(1)$ & $6(1)$ & $30(10)$ & $-$ & $<0.8$ \\ 
\multicolumn{3}{l}{\textbf{Photosensors \& Components}} &&&&&&&&&&&&&\\
\hline
18 & R11410 PMTs (average 180 PMTs) & Hamamatsu & Gator & - & - & mBq/PMT & $9(2)$ & $0.4(1)$ & $0.47(2)$ & $0.47(7)$ & $0.46(2)$ & $14.2(5)$ & $1.05(3)$ & $<0.14$ \\ 
19 & R11410 PMTs (average 60 PMTs) & Hamamatsu & GeMPI & - & - & mBq/PMT & $14(7)$ & $0.5(1)$ & $0.52(4)$ & $0.6(1)$ & $0.45(5)$ & $18.6(9)$ & $1.27(6)$ & $<0.13$ \\ 
20 & R11410 PMTs (average 99 PMTs) & Hamamatsu & GeMSE & - & - & mBq/PMT & $6.5(3)$ & $-$ & $0.32(4)$ & $0.33(5)$ & $0.19(1)$ & $11.1(4)$ & $0.71(3)$ & $<0.06$ \\ 
21 & Ceramic Stem & Hamamatsu & GeMPI & 1.5 & 20.7 & mBq/kg & $2.7(5)$ & $0.13(2)$ & $0.29(2)$ & $0.17(2)$ & $0.12(1)$ & $2.7(3)$ & $<0.003$ & $<0.009$ \\ 
22 & Ceramic Stem & Hamamatsu & GeMPI & 1.6 & 22.8 & mBq/kg & $3.4(5)$ & $0.12(2)$ & $0.22(1)$ & $0.20(2)$ & $0.07(1)$ & $0.13(2)$ & $<0.002$ & $<0.01$ \\ 
23 & Bases/components & Fralock/various & GeMSE & 1.9 & 7.0 & mBq/piece & $1.5(1)$ & $-$ & $0.7(3)$ & $0.14(1)$ & $0.053(3)$ & $0.29(5)$ & $<0.003$ & $<0.002$ \\ 
24 & HV Connectors/PTFE/Copper & Accu-Glass/custom & GeMSE & 1.0 & 27.5 & {\textmu}Bq/piece & $<310$ & $-$ & $<37$ & $<42$ & $<14$ & $140(90)$ & $50(10)$ & $<3.3$ \\ 
25 & Coaxial Cables & Huber \& Suhner & GeMPI & 1.5 & 22.6 & mBq/kg & $<74$ & $<1.4$ & $2.4(5)$ & $<1.5$ & $1.4(5)$ & $40(8)$ & $<0.2$ & $<0.6$ \\ 
26 & Kapton Cables & Accu-glass & GeMPI & 0.1 & 29.3 & mBq/kg & $<56$ & $<11$ & $8(2)$ & $<9.0$ & $<3.4$ & $110(30)$ & $<3.3$ & $<4.2$ \\ 
27 & Optical Fiber & Ratioplast & ICP-MS & $-$ & $-$ & mBq/kg & $7(3)$ & $-$ & $-$ & $7(2)$ & $-$ & $340(90)$ & $-$ & $-$ \\ 
\multicolumn{3}{l}{\textbf{Miscellaneous}} &&&&&&&&&&&&&\\
\hline
28 & M5 Screws (Ag), TPC & U-C Components & GeMPI & 0.5 & 27.6 & mBq/kg & $<43$ & $<1.4$ & $<1.1$ & $3(1)$ & $3.3(6)$ & $12(3)$ & $53(4)$ & $<0.6$ \\ 
29 & M6 Screws (Ag), TPC & U-C Components & GeMPI & 1.0 & 23.9 & mBq/kg & $<15$ & $<3.4$ & $<1.6$ & $<3.0$ & $2.7(7)$ & $18(5)$ & $6.7(7)$ & $<0.9$ \\ 
30 & M8 Screws (Ag), TPC & U-C Components & GeMPI & 0.9 & 22.8 & mBq/kg & $50(20)$ & $<2.1$ & $2.0(5)$ & $3.4(8)$ & $3.9(6)$ & $13(3)$ & $52(4)$ & $<0.8$ \\ 
31 & M8 Screws (Ag), Bell & ALCA & GeMPI & 1.2 & 21.6 & mBq/kg & $<29$ & $<0.8$ & $0.8(3)$ & $2.0(5)$ & $7.6(6)$ & $5(2)$ & $4.9(4)$ & $<0.3$ \\ 
32 & SMD Resistors & OHMITE & GeDSG & 0.01 & 6.9 & {\textmu}Bq/piece & $110(50)$ & $2.3(5)$ & $29(2)$ & $13(2)$ & $15(1)$ & $60(10)$ & $<1.1$ & $<0.6$ \\ 
\multicolumn{3}{l}{\textbf{Cleaning Solutions}} &&&&&&&&&&&&&\\
\hline
33 & HARO Clean 100 & HAROSOL & Corrado & 1.0 & 9.8 & mBq/kg & $<710$ & $<20$ & $11(6)$ & $<18$ & $<21$ & $4600(300)$ & $<8$ & $<7$ \\ 
34 & HARO Clean 188 & HAROSOL & Corrado & 1.5 & 6.0 & mBq/kg & $<24400$ & $-$ & $<35$ & $520 (310)$ & $<18$ & $5.8(4)\cdot 10^6$ & $<175$ & $<133$ \\ 
35 & HARO Clean 106 & HAROSOL & Corrado & 0.6 & 3.1 & mBq/kg & $2.1(1)\cdot 10^3$ & $-$ & $<20$ & $<77$ & $<16$ & $750(160)$ & $<9$ & $<15$ \\ 
36 & P3-Almeco 36 & Henkel & Bruno & 0.1 & 3.9 & mBq/kg & $14(3)\cdot 10^3$ & $790(160)$ & $98(27)$ & $<91$ & $<113$ & $15(2)\cdot 10^3$ & $<19$ & $<21$ \\ 
37 & Tickopur R33 & Dr. H. Stamm & Giove & 1.0 & 5.0 & mBq/kg & $<7400$ & $-$ & $36(13)$ & $<429$ & $<11$ & $1.55(9)\cdot 10^6$ & $<49$ & $<65$ \\ 
38 & Elma clean 65 & Elma & Giove & 1.0 & 7.1 & mBq/kg & $430(200)$ & $-$ & $6(3)$ & $<12$ & $<8$ & $1190(90)$ & $<1.6$ & $<4$ \\ 
39 & Elma clean 70 & Elma & Giove & 1.1 & 2.7 & mBq/kg & $<17100$ & $-$ & $53(18)$ & $<340$ & $19(9)$ & $1.45(9)\cdot 10^6$ & $<65$ & $<156$ \\ 
40 & HNO$_3$ (69\%) & Roth & Corrado & 0.5 & 5.1 & mBq/kg & $<970$ & $-$ & $<19$ & $<22$ & $<27$ & $<80$ & $<4$ & $<7$ \\ 
\end{tabular}
\caption{Measured activities of radioactive isotopes in the selected cryostat and TPC materials for XENONnT as well as the cleaning agents used during assembly. Measured values are given with $\pm1\sigma$ uncertainties, and upper limits are set at 95\% C.L. The facility, sample mass, and lifetime are provided for each measurement (as applicable). 
Samples 18-20 are the averaged results from measurements of different sets of 10-15 PMTs fitting into the germanium spectrometer at once.}
\label{tab:screening}
\end{table}
\end{landscape}

\thispagestyle{plain}

\begin{landscape}
\begin{table}[t]
\scriptsize
\vspace{3mm}
\begin{tabular}{l l l l l c c c c c c c c c c c}
\textbf{Sample} & \textbf{Component}& \textbf{Manufacturer}& \textbf{Facility}& \textbf{Mass [kg]}& \textbf{Livetime [d]}& \textbf{Units}& \textbf{$^{238}$U}& \textbf{$^{235}$U} & \textbf{$^{226}$Ra}& \textbf{$^{228}$Ra ($^{232}$Th)}& \textbf{$^{228}$Th}& \textbf{$^{40}$K}& \textbf{$^{60}$Co}& \textbf{$^{137}$Cs}\\
\hline 
\noalign{\vskip 2mm}
\multicolumn{3}{l}{\textbf{Neutron Veto}} &&&&&&&&&&&&&\\
\hline
41 & R5912 PMT Body & Hamamatsu & GeMPI & 0.2 & 6.9 & mBq/kg & $70(30)$ & $<6.4$ & $52(4)$ & $37(4)$ & $30(3)$ & $360(40)$ & $<1.8$ & $<1.3$ \\ 
42 & R5912 PMT low radioactivity Glass & Hamamatsu & GeCris & 0.4 & 3.9 & mBq/kg & $700(200)$ & $40(10)$ & $700(30)$ & $740(50)$ & $670(45)$ & $1000(100)$ & $<3.0$ & $<9.7$ \\ 
43 & Polyethylene PMT Holders & Plastotecnica emiliana & GeMPI & 0.1 & 13.8 & mBq/kg & $<19$ & $<2.2$ & $2.1(8)$ & $<2.2$ & $<2.1$ & $<22$ & $-$ & $<0.6$ \\ 
44 & SS Support Structure & Galli \& Morelli & GeMPI & 0.5 & 27.5 & mBq/kg & $<25$ & $<0.55$ & $0.8(2)$ & $1.1(4)$ & $2.3(4)$ & $<4.9$ & $4.1(4)$ & $<0.2$ \\ 
45 & ePTFE Reflectors & Applitecno Service & ICP-MS & - & - & mBq/kg & $0.3(1)$ & $-$ & $-$ & 0.12(4) & $-$ & $-$ & $-$ & $-$ \\ 
46 & Gadolinium Sulfate & NYC & GeMPI & 1.0 & 20.7 & mBq/kg & $<14$ & $<0.5$ & $0.9(2)$ & $0.4(2)$ & $1.2(2)$ & $<4.1$ & $-$ & $<0.06$ \\ 
47 & Gadolinium Sulfate & Treibacher & GeMPI & 1.0 & 6.6 & mBq/kg & $<43$ & $<3.6$ & $3.4(7)$ & $23(2)$ & $190(10)$ & $9(4)$ & $-$ & $<0.8$ \\ 
\multicolumn{3}{l}{\textbf{Purifying Getter Materials}} &&&&&&&&&&&&&\\
\hline
48 & LXe Filter 2$_{a}$ & BASF & Corrado & 0.3 & 23.8 & mBq/kg & $3200(700)$ & $-$ & $145(10)$ & $70(20)$ & $86(9)$ & $920(90)$ & $<6$ & $<7$ \\
49 & LXe Filter 2$_{b}$ & BASF & Corrado & 0.4 & 10.4 & mBq/kg & $2900(760)$ & $-$ & $150(10)$ & $70(20)$ & $84(16)$ & $720(90)$ & $<6$ & $<6$ \\
50 & LXe Filter 2$_{c}$ & BASF & Corrado & 0.1 & 7.2 & mBq/kg & $3600(1300)$ & $-$ & $1050(40)$ & $460(60)$ & $560(50)$ & $1500(200)$ & $<11$ & $<8$ \\
\multicolumn{3}{l}{\textbf{Calibration}} &&&&&&&&&&&&&\\
\hline
51 & Polyurethane Belt & BRECOFlex & GeMPI & 0.6 & 9.5 & mBq/kg & $46(19)$ & $<2.5$ & $13(1)$ & $5(1)$ & $4(1)$ & $93(17)$ & $1.9(6)$ & $<0.7$ \\ 
52 & Source Box & - & Giove & 7.3 & 13.8 & mBq/kg & $<26.4$ & $<1.7$ & $<0.9$ & $1.0(6)$ & $1.7(4)$ & $<1.2$ & $6.2(4)$ & $<0.2$ \\ 
53 & Source Box Clamp & McMaster & Corrado & 0.9 & 24.5 & mBq/kg & $<395$ & $<44$ & $46(4)$ & $<15.5$ & $15(5)$ & $<32$ & $12(2)$ & $<2$ \\ 
54 & SS316 Tube & Swagelok & GSOr & 2.9 & 7.1 & mBq/kg & $<33$ & $<2.8$ & $5.3(7)$ & $6.2(9)$ & $14(1)$ & $<7.5$ & $2.5(3)$ & $<0.8$ \\ 
55 & SS304 Beam Pipe & Weizmann & GeMPI & 3.1 & 16.7 & mBq/kg & $<63$ & $<1.8$ & $1.7(5)$ & $6(1)$ & $6(1)$ & $<10$ & $6.8(7)$ & $<0.5$ \\ 
56 & SS304 Support Pipe & Weizmann & GeMPI & 0.5 & 27.6 & mBq/kg & $36(18)$ & $<0.9$ & $0.6(3)$ & $1.9(8)$ & $6.8(8)$ & $<15$ & $6.5(8)$ & $<0.9$ \\ 
57 & SS316 Clamps & McMaster & GeMPI & 2.2 & 5.6 & mBq/kg & $150(60)$ & $<8.8$ & $102(5)$ & $19(2)$ & $18(2)$ & $7(3)$ & $12(1)$ & $-$ \\ 
58 & SS304 Bellow & MDC & GeMPI & 3.5 & 13.7 & mBq/kg & $<47$ & $<1.3$ & $2.0(4)$ & $5.3(9)$ & $9.8(9)$ & $<3.6$ & $7.1(7)$ & $<0.5$ \\ 
59 & SS304 Flanges & MDC & GeMPI & 12.0 & 20.6 & mBq/kg & $24(12)$ & $1.0(4)$ & $0.4(1)$ & $1.6(4)$ & $5.0(4)$ & $1.6(6)$ & $11.2(8)$ & $<0.2$ \\ 
60 & SS304 U-Tube & CG & GeMPI & 0.6 & 16.2 & mBq/kg & $<150$ & $<5.8$ & $<1.9$ & $<6.6$ & $4(1)$ & $15(8)$ & $45(4)$ & $<1.5$ \\
\end{tabular}
\caption{Measured activities of radioactive isotopes in the neutron veto, purification, and calibration subsystems. Measured values are given with $\pm1\sigma$ uncertainties, and upper limits are set at 95\% C.L. The facility, sample mass, and lifetime are provided for each measurement (as applicable). }
\label{tab:screening_other}
\end{table}

\begin{table}[t]
\scriptsize
\vspace{3mm}
\begin{tabular}{l l l l l c c c c c c c c c c c}
\textbf{Sample} & \textbf{Component}& \textbf{Manufacturer}& \textbf{Facility}& \textbf{Mass [kg]}& \textbf{Livetime [d]}& \textbf{Units}& \textbf{$^{238}$U}& \textbf{$^{235}$U} & \textbf{$^{226}$Ra}& \textbf{$^{228}$Ra ($^{232}$Th)}& \textbf{$^{228}$Th}& \textbf{$^{40}$K}& \textbf{$^{60}$Co}& \textbf{$^{137}$Cs}\\
\hline 
\noalign{\vskip 2mm}
R0 & SS304 \hspace{1.9cm} & Nironit \hspace{1.5cm} & GeMPI & 8.8 & 17.5 & mBq/kg & \hspace{0.20cm} $<21$ \hspace{0.20cm} & \hspace{0.20cm} $<0.8$ \hspace{0.20cm}& \hspace{0.20cm} $1.7(2)$\hspace{0.20cm} & \hspace{0.20cm} $1.1(3)$ \hspace{0.20cm}& \hspace{0.20cm} $1.2(2)$ \hspace{0.20cm}& \hspace{0.20cm} $<3.0$ \hspace{0.20cm}& \hspace{0.20cm} $0.8(1)$ \hspace{0.20cm}& \hspace{0.20cm} $<0.2$\hspace{0.20cm}  \\ 
R0 & SS304 & Nironit & ICP-MS & $-$ & $-$ & mBq/kg & $3.7(5)$ & $-$ & $-$ & $0.5(1)$ & $-$ & $-$ & $-$ & $-$ \\ 
R1 & SS20 & Terni & GeMPI & 10.3 & 12.3 & mBq/kg & $<24$ & $<1.2$ & $<69$ & $2.1(4)$ & $7.1(6)$ & $1.3(6)$ & $14(1)$ & $<0.2$ \\ 
R1 & SS20 & Terni & ICP-MS & $-$ & $-$ & mBq/kg & $4.5(8)$ & $-$ & $-$ & $5.0(8)$ & $-$ & $-$ & $-$ & $-$ \\ 
R2 & SS27 & Terni & GeMPI & 7.2 & 16.6 & mBq/kg & $<40$ & $<1.4$ & $<0.6$ & $<1.1$ & $2.1(4)$ & $<1.4$ & $23(2)$ & $<0.2$ \\ 
R2 & SS27 & Terni & ICP-MS & $-$ & $-$ & mBq/kg & $15(3)$ & $-$ & $-$ & $1.4(2)$ & $-$ & $-$ & $-$ & $-$ \\ 
R3 & SS304 & Nironit & GeMPI & 8.0 & 13.8 & mBq/kg & $<11$ & $<0.9$ & $2.0(2)$ & $2.2(3)$ & $5.4(4)$ & $1.7(5)$ & $0.31(5)$ & $<0.1$ \\ 
R3 & SS304 & Nironit & ICP-MS & $-$ & $-$ & mBq/kg & $3.5(6)$ & $-$ & $-$ & $3(1)$ & $-$ & $-$ & $-$ & $-$ \\ 
\end{tabular}
\caption{Measured activities of radioactive isotopes in some materials that were rejected. Measured values are given with $\pm1\sigma$ uncertainties, and upper limits are set at 95\% C.L. The facility, sample mass, and lifetime are provided for each measurement (as applicable).}
\label{tab:screening_rejected}
\end{table}
\end{landscape}



\thispagestyle{plain}

\begin{landscape}
\begin{table}[t]
\scriptsize
\begin{tabular}{l l l l l l c c r r l}
\textbf{Sample} & \textbf{Component} & \textbf{Manufacturer}& \textbf{Description} & \textbf{Facility}& \textbf{Mass [kg]}&\textbf{Length [m]}& \textbf{Units}& \textbf{$^{222}$Rn} & \multicolumn{2}{c}{\textbf{Used in XENONnT}}\\
\hline 
\noalign{\vskip 3mm}
\multicolumn{3}{l}{\textbf{Cables in CRY}} &&&&&&& {\scriptsize in Cryostat [m]} & {\scriptsize in Cable Pipe 2 [m]} \\
\hline
\rnAccuglassa & Accuglass-1 & Reliable Hermetic Seals  & Accuglass 100670 single-core kapton wire & PC & 0.071 & 100 & mBq/m &  $\le$0.28\,$\cdot$\,10$^{-3}$& 0&0 \\
\rnAccuglassb & Accuglass-2 & Reliable Hermetic Seals & Accuglass 100670 single-core kapton wire & PC & 0.070 & 100 & mBq/m & 0.55\,(19)\,$\cdot$\,10$^{-3}$&  0 & 998 \\
\rnAccuglassc & Accuglass-3 & Reliable Hermetic Seals & Accuglass 100670 single-core kapton wire & PC & $-$ & 200 & mBq/m & 0.41\,(15)$\cdot$10$^{-3}$ & 1821 & 0  \\
\rnHabia & Habia RG196A/U & Koax24 & PTFE coaxial cable RG196 & PC & 0.999 & 100 & mBq/m & 12\,(2)\,$\cdot$\,10$^{-3}$& 0&0 \\
\rnHubersuhnera & H+S 1 & Novitronic AG & Huber+Suhner RG196 A/U & PC & $-$ & 50  & mBq/m  & $<$\,0.48\,$\cdot$\,10$^{-3}$&   192 & 442\\
\rnHubersuhnerb & H+S 2 & Huber+Suhner & Huber+Suhner RG196 A/U & PC & $-$ & 100 & mBq/m & 0.38\,(22)\,$\cdot$\,10$^{-3}$&  991 & 0  \\
\rnHubersuhnerc & H+S 3 & Huber+Suhner & Huber+Suhner RG196 A/U & PC & $-$ & 50 & mBq/m  & $<$\,1.36\,$\cdot$\,10$^{-3}$&  264 & 614\\
\rnPasternack & Pasternack & Pasternack E. & Pasternack RG196 A/U  & PC & $-$ & 198 & mBq/m & 0.48\,(18)\,$\cdot$\,10$^{-3}$& 0&0 \\
\noalign{\vskip 3mm}
\multicolumn{3}{l}{\textbf{Plastics in Rn-DST}} &&&&&&&  \\
\hline
\rnUhmwpe & UHMWPE & $-$ & gasket material & PC & 0.160 & $-$ & mBq & 0.09\,(2) & \multicolumn{2}{c}{ Mag-Pump 4 }  \\
\rnIglidurwhite & Iglidur A180 & Igus & gasket material white & PC & 0.178 & $-$ & mBq & $\le$\,0.05 & \multicolumn{2}{c}{ Mag-Pump 3}\\
\rnIgliduryellow & Iglidur J & Igus & gasket material yellow & PC & 0.266 & $-$ & mBq & $\le$\,0.02 & \multicolumn{2}{c}{ Mag-Pump 1 \& 2}\\
\rnIglidurblue & Iglidur A350 & Igus & gasket material blue & PC & 0.145 & $-$ & mBq & 0.06\,(3) & \multicolumn{2}{c}{NO} \\
\noalign{\vskip 3mm}
\multicolumn{3}{l}{\textbf{LXe pump \& filter in LXe-PUR}} &&&&&& \\
\hline
\rnViton & Viton O-ring & Barber Nichols & 0.103$\times$4.487 fitting & PC & 0.004  & $-$ & mBq & 0.76\,(8) & \multicolumn{2}{c}{YES} \\
\rnSSvacuumline & SS cyogenic valve & Thermomess & bellows-sealed globe valve (after installation)  & PC & $-$ & $-$ & mBq/valve & 0.22\,(4) & \multicolumn{2}{c}{YES} \\
\rnSScryovalve & SS cyogenic valve & Thermomess & bellows-sealed globe valve & PC & 2.713 & $-$ & mBq/valve & 0.11\,(4)  & \multicolumn{2}{c}{YES} \\
\rnAlrotor & Aluminum rotor & Barber Nichols & Alumina rotor from LXe-Pump  & PC & 0.062 & 0.06 & mBq & $<$\,0.017 & \multicolumn{2}{c}{YES} \\
\rnXenTLXeFilterQa & LXe Filter 2$_{a}$ & BASF  & Sample 48 in Table~\ref{tab:screening_other} & PC & 0.3 & $-$ & mBq/kg & 50\,(2) & \multicolumn{2}{c}{NO}  \\
\rnXenTLXeFilterQb & LXe Filter 2$_{c}$ & BASF & Sample 50 in Table~\ref{tab:screening_other} & PC & 0.271 & $-$ & mBq/kg & 369\,(15) & \multicolumn{2}{c}{NO}  \\
\noalign{\vskip 3mm}
\end{tabular}
 \caption{
 \label{tab:rn_items} 
 $^{222}$Rn emanation rate investigation of materials and components from XENONnT subsystems.
 Columns indicate the sample ID; sample name, manufacturer and description; and the measurement facility used: proportional counter (PC).
 Mass and/or length are provided when applicable.
 Measured values are given with $\pm1\sigma$ uncertainties and upper limits are set at 90\% C.L.
}
\end{table}
\end{landscape}

\thispagestyle{plain}
\begin{landscape}
\begin{table}[t]
\scriptsize

\centering
\begin{tabular}{l l ll c }
\textbf{Sample} & \textbf{Component} & \textbf{Description and/or manufacturer} &  \textbf{Facility} & \textbf{$^{222}$Rn [mBq]} \\
\hline
\noalign{\vskip 3mm}
\multicolumn{3}{l}{\textbf{CRY}} \\
\hline
\rnXenTCryostat & Cryostat  & Costruzioni Generalli  & PC & 1.9\,(2) \\
\rnXenTCryogenics & Cooling Tower, Cryostat Pipe, HE \& Heater, & retained from XENON1T & RM & 17\,(3)  \\
 &  Cable Feedthrough 1 \& Cable Pipe 1 &  &  &   \\
\rnXenTNewCableFeedthrough & Cable Feedthrough 2 & $-$  &  PC & 1.6\,(2)   \\
\rnXenTNewCablePipe & Cable Pipe 2 & $-$ & PC & 0.9\,(1)  \\
$-$ & \textbf{ Integral CRY} & $-$ & $-$ & \textbf{\rnXenTCryoIntegral}\\ 
\noalign{\vskip 1mm}
\multicolumn{3}{l}{\textbf{GXe-PUR}} \\
\hline
\rnXenTGXeMaga\, in~\cite{ref:xenon1t_radon} & Mag-pump 1 & Muenster U. & $-$ & 0.3\,(1)  \\
\rnXenTGXeGettera\, in~\cite{ref:xenon1t_radon} & GXe Filter 1 & SAES ($\sim$4 kg) & $-$ & 1.2\,(2) \\
\rnXenTGXeGetterb\, in~\cite{ref:xenon1t_radon} & GXe Filter 2 & SAES ($\sim$4 kg) & $-$ & 0.24\,(3) \\
$-$ & \textbf{ Integral GXe-PUR} & $-$ & $-$ & \textbf{\rnXenTGXeIntegral}\\ 
\noalign{\vskip 1mm}
\multicolumn{3}{l}{\textbf{LXe-PUR}} \\
\hline
\rnXenTLXeVacumPipe & Vacuum insulated pipe & Costruzioni Generalli & PC & 0.22\,(5) \\
\rnXenTLXePurityMonitor & Purity Monitor & U. Tokyo & PC & 0.25\,(6)  \\
\rnXenTLXeFilterST & LXe Filter 1 & SAES St707 ($\sim$3 kg) & PC & 0.24\,(3) \\
\rnXenTLXeHeatExchanger & Heat exchanger & DATE, 500W Xe/LN$_{2}$    & PC & 1.3\,(1)  \\
\rnXenTLXePumpb & LXe-Pump  & Barber Nichols & PC & 1.6\,(2) \\ 
$-$ & \textbf{ Integral LXe-PUR} & $-$ & $-$ & \textbf{\rnXenTLXeIntegral}\\ 
\noalign{\vskip 1mm}
\multicolumn{3}{l}{\textbf{Rn-DST}} \\
\hline
\rnXenTRADColumn & Distillation Column & $-$ & PC & \multicolumn{1}{c}{1.7\,(2)} \\
 \#20$_{\textrm{a}}$ in~\cite{ref:xenon1t_radon} & Q-Drive  & $-$ & $-$ & \multicolumn{1}{c}{2.5\,(1)} \\
\rnXeoneTRADgetter & GXe Filter 3 &  SAES ($\sim$0.5 kg) & estimated from \#18 in~\cite{ref:xenon1t_radon}& \multicolumn{1}{c}{0.09\,(3)}  \\
\rnXenTRADMagcdef & 4 Mag-pumps & $-$ & PC & \multicolumn{1}{c}{0.30\,(5)} \\
\rnXenTRADbufferin  & Mag-Pumps inlet buffer &  U. Muenster  & estimated from~\cite{ref:xenon1t_radon,ref:galex_prop_counters} & \multicolumn{1}{c}{0.24\,(12)} \\
\rnXenTRADbufferout & Mag-Pumps outlet buffer &  Hositrad & PC &   \multicolumn{1}{c}{0.94\,(9)}\\
\rnXenTRADSecespol & Xe-Xe heat exchanger & Secespol  & PC &  \multicolumn{1}{c}{$<$\,0.03} \\
\rnXenTRADreboilerdown & Xe condenser & U. Muenster & estimated from~\cite{ref:xenon1t_radon,ref:galex_prop_counters} & \multicolumn{1}{c}{0.07\,(4)} \\
$-$ & \textbf{ Integral Rn-DST (\rnXenTRADMagcdef-\rnXenTRADreboilerdown)} & $-$ & $-$ & \textbf{\rnXenTRnDSTIntegral}\\
\noalign{\vskip 1mm}
\multicolumn{3}{l}{\textbf{TPC}} \\
\hline
\rnXenTtpc & \textbf{ Integral TPC} & $-$ & RM & \textbf{ \rnXenTtpcIntegral }  \\
\end{tabular}
\caption{\label{tab:rn_XENONnT_samples}
$^{222}$Rn emanation rate from xenon handling systems and the TPC of XENONnT.
Columns indicate the measurement ID; sample name, manufacturer and description; and the measurement facility used: proportional counter or radon monitor (PC or RM).
Previous results of XENON1T are further referenced with their ID in~\cite{ref:xenon1t_radon}.
Measured values are given with $\pm1\sigma$ uncertainties and upper limits are set at 90\%~C.L.
}
\end{table}
\end{landscape}

\end{document}

%% file: authorlist_20210921.tex





\author{E.~Aprile\thanksref{addr3}
\and
K.~Abe\thanksref{addr24}
\and
F.~Agostini\thanksref{addr0}
\and
S.~Ahmed Maouloud\thanksref{addr18}
\and
M.~Alfonsi\thanksref{addr5}
\and
L.~Althueser\thanksref{addr7}
\and
E.~Angelino\thanksref{addr14}
\and
J.~R.~Angevaare\thanksref{addr8}
\and
V.~C.~Antochi\thanksref{addr12}
\and
D.~Ant\'on Martin\thanksref{addr1}
\and
F.~Arneodo\thanksref{addr9}
\and
L.~Baudis\thanksref{addr17}
\and
A.~L.~Baxter\thanksref{addr10}
\and
L.~Bellagamba\thanksref{addr0}
\and
R.~Biondi\thanksref{addr4}
\and
A.~Bismark\thanksref{addr17}
\and
A.~Brown\thanksref{addr19}
\and
S.~Bruenner\thanksref{addr8,addr6,corrauthors}
\and
G.~Bruno\thanksref{addr9,addr13}
\and
R.~Budnik\thanksref{addr16}
\and
C.~Capelli\thanksref{addr17}
\and
J.~M.~R.~Cardoso\thanksref{addr2}
\and
D.~Cichon\thanksref{addr6}
\and
B.~Cimmino\thanksref{addr21}
\and
M.~Clark\thanksref{addr10}
\and
A.~P.~Colijn\thanksref{addr8,addr31}
\and
J.~Conrad\thanksref{addr12}
\and
J.~J.~Cuenca-Garc\'ia\thanksref{addr27}
\and
J.~P.~Cussonneau\thanksref{addr13}
% ...
(XENON Collaboration) 
\and\\
M.~Laubenstein\thanksref{addr4}
\and
S.~Nisi\thanksref{addr4}.
}
\newcommand{\bologna}{Department of Physics and Astronomy, University of Bologna and INFN-Bologna, 40126 Bologna, Italy}
\newcommand{\chicago}{Department of Physics \& Kavli Institute for Cosmological Physics, University of Chicago, Chicago, IL 60637, USA}
\newcommand{\coimbra}{LIBPhys, Department of Physics, University of Coimbra, 3004-516 Coimbra, Portugal}
\newcommand{\columbia}{Physics Department, Columbia University, New York, NY 10027, USA}
\newcommand{\lngs}{INFN-Laboratori Nazionali del Gran Sasso and Gran Sasso Science Institute, 67100 L'Aquila, Italy}
\newcommand{\mainz}{Institut f\"ur Physik \& Exzellenzcluster PRISMA$^{+}$, Johannes Gutenberg-Universit\"at Mainz, 55099 Mainz, Germany}
\newcommand{\heidelberg}{Max-Planck-Institut f\"ur Kernphysik, 69117 Heidelberg, Germany}
\newcommand{\munster}{Institut f\"ur Kernphysik, Westf\"alische Wilhelms-Universit\"at M\"unster, 48149 M\"unster, Germany}
\newcommand{\nikhef}{Nikhef and the University of Amsterdam, Science Park, 1098XG Amsterdam, Netherlands}
\newcommand{\nyuad}{New York University Abu Dhabi - Center for Astro, Particle and Planetary Physics, Abu Dhabi, United Arab Emirates}
\newcommand{\purdue}{Department of Physics and Astronomy, Purdue University, West Lafayette, IN 47907, USA}
\newcommand{\rice}{Department of Physics and Astronomy, Rice University, Houston, TX 77005, USA}
\newcommand{\stockholm}{Oskar Klein Centre, Department of Physics, Stockholm University, AlbaNova, Stockholm SE-10691, Sweden}
\newcommand{\subatech}{SUBATECH, IMT Atlantique, CNRS/IN2P3, Universit\'e de Nantes, Nantes 44307, France}
\newcommand{\torino}{INAF-Astrophysical Observatory of Torino, Department of Physics, University  of  Torino and  INFN-Torino,  10125  Torino,  Italy}
\newcommand{\ucsd}{Department of Physics, University of California San Diego, La Jolla, CA 92093, USA}
\newcommand{\wis}{Department of Particle Physics and Astrophysics, Weizmann Institute of Science, Rehovot 7610001, Israel}
\newcommand{\zurich}{Physik-Institut, University of Z\"urich, 8057  Z\"urich, Switzerland}
\newcommand{\paris}{LPNHE, Sorbonne Universit\'{e}, Universit\'{e} de Paris, CNRS/IN2P3, 75005 Paris, France}
\newcommand{\freiburg}{Physikalisches Institut, Universit\"at Freiburg, 79104 Freiburg, Germany}
\newcommand{\lal}{Universit\'{e} Paris-Saclay, CNRS/IN2P3, IJCLab, 91405 Orsay, France}
\newcommand{\napels}{Department of Physics ``Ettore Pancini'', University of Napoli and INFN-Napoli, 80126 Napoli, Italy}
\newcommand{\nagoya}{Kobayashi-Maskawa Institute for the Origin of Particles and the Universe, and Institute for Space-Earth Environmental Research, Nagoya University, Furo-cho, Chikusa-ku, Nagoya, Aichi 464-8602, Japan}
\newcommand{\laquila}{Department of Physics and Chemistry, University of L'Aquila, 67100 L'Aquila, Italy}
\newcommand{\tokyo}{Kamioka Observatory, Institute for Cosmic Ray Research, and Kavli Institute for the Physics and Mathematics of the Universe (WPI), University of Tokyo, Higashi-Mozumi, Kamioka, Hida, Gifu 506-1205, Japan}
\newcommand{\kobe}{Department of Physics, Kobe University, Kobe, Hyogo 657-8501, Japan}
\newcommand{\ucla}{Physics \& Astronomy Department, University of California, Los Angeles, CA 90095, USA}
\newcommand{\kit}{Institute for Astroparticle Physics, Karlsruhe Institute of Technology, 76021 Karlsruhe, Germany}
\newcommand{\tsinghua}{Department of Physics \& Center for High Energy Physics, Tsinghua University, Beijing 100084, China}
\newcommand{\alsoatferrara}{INFN, Sez. di Ferrara and Dip. di Fisica e Scienze della Terra, Universit\`a di Ferrara, via G. Saragat 1, Edificio C, I-44122 Ferrara (FE), Italy}
\newcommand{\alsoatsuny}{Simons Center for Geometry and Physics and C. N. Yang Institute for Theoretical Physics, SUNY, Stony Brook, NY, USA}
\newcommand{\alsoatutrecht}{Institute for Subatomic Physics, Utrecht University, Utrecht, Netherlands}
\newcommand{\alsoatcoimbrapoli}{Coimbra Polytechnic - ISEC, 3030-199 Coimbra, Portugal}
\newcommand{\alsoatiarnagoya}{Institute for Advanced Research, Nagoya University, Nagoya, Aichi 464-8601, Japan}
\authorrunning{XENON Collaboration}

\thankstext{corrauthors}{Corresponding author: {\color{blue}stefanb@nikhef.nl}, {\color{blue}jpalacio@mpi-hd.mpg.de}, {\color{blue}shayne@physik.uzh.ch}, {\color{blue}xenon@lngs.infn.it}}
\thankstext{addr31}{Also at \alsoatutrecht}
\thankstext{addr33}{Also at \alsoatiarnagoya}
\thankstext{addr32}{Also at \alsoatcoimbrapoli}
\thankstext{addr29}{Also at \alsoatferrara}


\institute{\columbia\label{addr3}
\and
\tokyo\label{addr24}
\and
\bologna\label{addr0}
\and
\paris\label{addr18}
\and
\mainz\label{addr5}
\and
\munster\label{addr7}
\and
\torino\label{addr14}
\and
\nikhef\label{addr8}
\and
\stockholm\label{addr12}
\and
\chicago\label{addr1}
\and
\nyuad\label{addr9}
\and
\zurich\label{addr17}
\and
\purdue\label{addr10}
\and
\lngs\label{addr4}
\and
\freiburg\label{addr19}
\and
\heidelberg\label{addr6}
\and
\subatech\label{addr13}
\and
\wis\label{addr16}
\and
\coimbra\label{addr2}
\and
\napels\label{addr21}
\and
\kit\label{addr27}
\and
\rice\label{addr11}
\and
\laquila\label{addr23}
\and
\tsinghua\label{addr28}
\and
\nagoya\label{addr22}
\and
\ucsd\label{addr15}
\and
\lal\label{addr20}
\and
\kobe\label{addr25}
}